# LATE-TYPE STARS IN M31: I. PHOTOMETRIC STUDY OF AGB STARS AND METALLICITY GRADIENTS

James P. Brewer[1], Harvey B. Richer
Department of Geophysics and Astronomy, University of British Columbia, 129-2219 Main Mall, Vancouver, B.C., V6T 1Z4, Canada
Electronic mail: brewer@astro.ubc.ca, richer@astro.ubc.ca

Dennis R. Crabtree[1]
Dominion Astrophysical Observatory, Herzberg Institute of Astrophysics, National Research Council, 5071 W. Saanich Road, Victoria, B.C., V8X 4M6, Canada
Electronic mail: crabtree@dao.nrc.ca

## ABSTRACT

We have imaged five $7' \times 7'$ fields in M31 spanning galactocentric radii from 4 to 32 kpc along the SW-major axis. The fields were observed through two broad-band ($V$ and $I$) and two narrow-band ($CN$ and $TiO$) filters. The broad-band data were used to construct $I$, $(V-I)$ color-magnitude diagrams (CMDs) and, in some of our fields, we found significant numbers of stars in the Cepheid instability strip. A distance modulus for the Cepheids in the middle field was found that agreed well with other values in the literature values. The width of the giant branch (GB) in the $I$, $(V-I)$ CMD of all 5 fields was investigated, and we show that in four of the fields a likely explanation for the GB width is a combination of *both* metallicity and mass variations. Using the broad-band data, the asymptotic giant branch (AGB) luminosity functions (LFs) were measured in the five fields, and we show that differences exist between these LFs. We speculate on how the different star forming histories in the fields may lead to the observed AGB LFs and GB widths. Using the narrow-band data along with the broad-band data we separated the AGB stars into carbon-rich (C) and oxygen-rich (M) types. The carbon stars LFs were used to obtain an estimate for the distance modulus of M31 which agrees with the value derived from Cepheids. The ratio of C- to M-stars (C/M) is believed to be an indicator of gaseous chemical abundance at the time of formation of these stars. We show that the C/M ratio increases smoothly with galactocentric distance, suggesting an inverse correlation with metallicity. This is the first demonstration of this effect within a single extragalactic system. We find that differences in the width of the GB and the AGB LFs do not significantly affect the C/M ratio. We consider the effect of the increasing C/M ratio on the ISM in M31, and cite evidence in favor of a model where the grain composition in M31 is a function of galactocentric distance.

[1]Visiting Astronomer, Canada–France–Hawai'i Telescope (CFHT), operated by the National Research Council of Canada, le Centre National de la Recherche Scientifique de France, and the University of Hawai'i.

## 1. INTRODUCTION

Understanding the color and chemical evolution of galaxies requires knowledge of the asymptotic giant branch (AGB). The AGB is the luminous final phase of stellar evolution for low- and medium-mass stars ($0.8\mathcal{M}_\odot \lesssim M \lesssim 10\mathcal{M}_\odot$) and is the only way to study these previously faint stars in other galaxies. This in turn allows us to investigate the stellar populations of such systems. As a star evolves up the AGB it is burning both hydrogen and helium in shells. The double shell source was shown by Schwarzschild & Härm (1965) to be thermally unstable. With improved computing power, Iben (1975) was later able to show that the thermal instabilities can lead to a process known as "third dredge-up" in which processed material gets deposited in the stellar envelope. The third dredge-up can alter the surface composition of an M-star (C/O <1) on the AGB, transforming it to either an S-star (C/O≃1) or a carbon star (C-star, C/O >1). In this paper, unless otherwise stated, the spectral types M, S, and C refer to AGB stars.

Stars on the AGB can make a significant contribution to the integrated light of a galaxy. Renzini (1981) estimated that the AGB contributes approximately 50% to the integrated luminosity of a system which is approximately $2 \times 10^8$ years old (although in a system as old as a Galactic globular cluster this contribution will have dropped to around 10%). At cosmological distances we only observe the integrated color and spectrum of a galaxy, and to understand these observations we need a full understanding of the AGB. For example, incorporating AGB evolution into galactic models allowed Lilly (1987) to explain the locus of galaxies in the ($U-V$, $V-H$) plane.

A recent review of mass loss from AGB stars is given by Knapp (1991). It appears that AGB stars are losing mass at a rate of $10^{-8}$ to $10^{-4}\mathcal{M}_\odot$ yr$^{-1}$. Indeed, the high mass-loss rate of C-stars and their low temperatures have caused them to be likened to smoking candles. Knapp (1991) states that it appears that AGB stars can account for the bulk of material returned to the interstellar medium (ISM) by stellar evolutionary processes. This material contributes to the radiative cooling of dense interstellar clouds, ending in their possible collapse and subsequently star formation. Due to its high mass-loss rate, an AGB star can avoid becoming a supernova by ejecting its outer envelope before its core mass grows above the Chandrasekhar limit. The fate of a star (planetary nebula or supernova) has strong consequences for both the composition and the blast-wave energy input of the ISM.

Aaronson *et al.* (1989) discussed the properties of C-stars which make them useful as probes of Galactic structure. They mention the observational evidence of a narrow C-star luminosity function, the high luminosity of C-stars which allows them to be observed at the edge of the Galaxy through several magnitudes of absorption, and the fact that they are relatively abundant. Also discussed by Aaronson *et al.* (1989) are: the ease with which C-stars can be detected at low Galactic latitudes on objective prism plates; their rich spectra allowing velocity determinations from low signal-to-noise exposures and their age (1-5 Gyrs), which means they have lost memory of their formation velocities.

To take full advantage of these features of C-stars as possible probes of Galactic structure, we must quantify any dependence (or lack thereof) their luminosity function has on metallicity. Comprehensive study of C-stars in a nearby spiral like M31 is a step toward this goal.

To further the knowledge of the AGB, groups led by Richer (Richer, Crabtree & Pritchet 1984 (henceforth RCP84), Richer & Crabtree 1985 (henceforth RC85), Richer, Pritchet & Crabtree 1985, Pritchet *et al.* 1987, Hudon *et al.* 1989, and Richer, Crabtree & Pritchet 1990 (henceforth RCP90)) and the late Marc Aaronson (Aaronson *et al.* 1984, Cook, Aaronson & Norris 1986) developed and refined a technique, similar to one originally suggested by Palmer & Wing (1982), to identify C- and M-stars in crowded fields. The technique involves imaging a field through four filters. Two narrow band filters provide low-resolution spectral information on CN and TiO band strengths, while the $(V-I)$ color provides a temperature discriminator. The strength of the photometric technique is that it allows surveys to be made in fields otherwise too crowded for grism or grens methods, while still offering a multiplex advantage unlike spectroscopy of individual stars. Use of this method has pushed surveys of AGB stars beyond the Milky Way and its dwarf companions; the most distant C-stars currently known, with a true distance modulus of 27.4, were identified by Hudon *et al.* (1989) in NGC 2403 using this technique.

In combination with the above photometric system, charge-coupled devices (CCDs),

with their high quantum efficiency and linearity, make ideal detectors for surveys of AGB stars in external systems. The small field size of a CCD will still enable a significant fraction of a distant galaxy to be observed, while the data from a CCD can be reduced in a standard manner using software that allows accurate photometry to be obtained in crowded fields. A three band version of the system has recently been used by Green *et al.* (1994) to search for high latitude dwarf C-stars, using CCD detectors. We point out that in uncrowded fields the system will work with photographic plates and plate scanners, and despite the disadvantages of these over CCDs, the use of photographic plates has the potential to greatly increase sky coverage.

The aims of this investigation are to identify and classify AGB stars along the major-axis of M31. This information can then be used to study the interplay between AGB properties, such as the C-star luminosity function or the ratio of C- to M-stars, and differences in the local properties of M31, such as metallicity or star-forming history. By investigating the interplay between AGB and local properties in a single system we eliminate the need to take into consideration the effects of galactic morphological type.

An ideal candidate for such a study is M31; it is sufficiently close that its AGB stars can be easily measured with a four-meter class telescope while it is sufficiently distant that a significant fraction of it can be observed in a single CCD field. An abundance gradient has been previously observed along the major-axis of M31, and our observations were chosen so that we have differing star-forming histories in our fields.

## 2. OBSERVATIONS, REDUCTIONS, AND CALIBRATION

### 2.1. Observations

Our observations were made at the CFHT with the FOCAM camera at the prime focus. The detector was a Lick2 2048×2048 CCD which is a thick CCD coated with a laser dye to improve the blue efficiency. The CCD is cosmetically good and has a quantum efficiency which rises from 30% to 60% between 5000Å and 8000Å. The pixel size on the CCD is 15 $\mu m^2$ which gives a plate scale of 0$''$.205/pixel and a field size of approximately $7' \times 7'$.

The five fields observed were all located along the SW-major axis of M31. The locations of the fields are easily identified using features shown in Hodge's (1981) *Atlas of the Andromeda Galaxy.* Field 1, the innermost field, is approximately centered on the dust cloud D152 while Fields 2, 3 and 4 are roughly centered on the dust cloud D61 and the globular clusters G53 and G31 respectively. No distinctive features are located in Field 5; the center of this field has coordinates of $\alpha = 0^h33^m00^s$, $\delta = 39°12'.5$ (equinox 1950.0) on Hodge's (1981) Chart 16. Coordinates (equinox 2000.0) for all 5 field centers are given in Table 1, while binned $I$ band images of each field are on the AAS CD-ROM Series, volume XX, 1995. The approximate sizes, orientations and locations of the five M31 fields are shown in Fig. 1 which is a montage of 2 plates, E-398 and E-851, from the Palomar Sky Survey. Figure 1 gives a good indication of the fraction of M31 that was imaged together with the radial extent of the fields.

In Table 2 we present a list of stars from the Guide Star Catalog[2], variable stars (Baade & Swope 1965), and clusters in M31 (Hodge 1981) that we reidentified on our fields. The version of Table 2 in this paper is for guidance regarding form and content only, Table 2 is presented in its complete form on the AAS CD-ROM Series, volume XX, 1995. The first column heading in Table 2 gives the field in which the star/cluster was identified, the second column gives the X,Y coordinates which can be used to identify the star/cluster using the appropriate $I$-band image on the same volume of the AAS CD-ROM as the complete version of Table 2. Column 3 identifies the type of star/cluster, while column 4 gives an ID from the literature and column 5 gives the reference for this ID.

The fields were observed through filters that approximated the Johnson $V$ and Cousins $I$ system, and through two narrow-band filters ($\Delta\lambda \sim 140$Å) which were centered on the CN ($\Delta\nu = +2$, 8100Å) and TiO ($\Delta\nu = -1$, 7800Å) bands, henceforth the $CN$ and $TiO$ filters. Integration times for the broad- and narrow-band images were 600 and 1200 seconds respectively, with three or more frames being obtained in each filter in each field. The seeing conditions throughout the four night CFHT run were superb; the mean seeing on the combined frames was 0\".76, and all but the last night were deemed photometric. A condensed observing log is given in Table 1

### 2.2. Reductions

Using standard routines in IRAF [3] the raw frames were bias-subtracted and divided by flat fields which were made from images of the sky and the inside of the dome. Low levels of vignetting were apparent in the corners of some frames, the level of which was small enough that it could be ignored. Due to crowding, the $CN$, $TiO$, $V$, and $I$ frames for each field were made by combining only those frames with the best seeing. The frames were averaged together using the IMCOMBINE task in IRAF, with either AVSIGCLIP or CRREJECT used to eliminate charged-particle events. The total exposure and resultant full-width at half-maximum (FWHM) of the stellar images on each (combined) frame is given in Table 1.

---

[2]The GSC was prepared by the STScI operated by the Association of Universities for Research in Astronomy, Inc., for the National Aeronautics and Space Administration.

[3]Image Reduction and Analysis Facility (IRAF), a software system distributed by the National Optical Astronomy Observatories (NOAO).

All subsequent photometric measurements on the combined frames were made using the photometry reduction packages DAOPHOT II and ALLSTAR (Stetson *et al.* 1990; Stetson 1992), both of which have been extensively discussed in the literature. After building a point-spread function (PSF) from typically 20 to 30 stars on each frame, we used a "double pass" of FIND/PHOT/ALLSTAR to reduce our frames. Briefly, the steps involved in a double pass are: (1) search the frame for stars using a finding threshold of $4\sigma$ (standard deviation of the sky noise); (2) use the frame's PSF along with ALLSTAR to produce a frame on which the stars from (1) have been subtracted out; (3) search the subtracted frame (finding threshold of $4\sigma$) for stars which had been missed in (1); (4) combine the list of subtracted stars with the list of stars from (3); and finally (4) rereduce the original frame using ALLSTAR and the combined list of stars. Due to the CCD frame size (2K$\times$2K) and the large number of stars found on some frames, it was necessary to increase the dimensions of certain arrays in DAOPHOT II and ALLSTAR. The number of stars fitted on each of the 20 combined frames is given in column 5 of Table 1. To match the stars between the $CN$, $TiO$, $V$, and $I$ frames of each field we used DAOMASTER (Stetson 1993), saving only those stars that were found on at least two of the four frames.

### 2.3. Calibration

Standards were observed through $V$ and $I$ filters on nights 1 and 2 of the run and a summary of these observations can be found in Table 3. The NGC 2419 and NGC 7006 frames were reduced with a double pass (see Sec. 2.2). All the non-standard stars were subtracted leaving only the standards on the frame. The aperture magnitudes of all the standards were measured using DAOGROW (Stetson 1990), and were corrected for exposure time and airmass. Extinction coefficients for $V$ ($Q_y$) and $(V-I)$ ($k_9$) were derived from observations of NGC 7006 at various airmasses and found to be $Q_y = 0.40 \pm 0.05$ and $k_9 = 0.17 \pm 0.06$ respectively. These higher than normal values can be attributed to the eruption of Mt. Pinatubo in the Philippines which occurred 3 months prior to our observations. It is known that volcanic dust can cause extinction coefficients to vary by over a factor of 3 in the optical (Krisciunas 1994). The transformation equations between aperture magnitudes and standard magnitudes had color terms consistent with zero, indicating that both the $V$ and $I$ filters were good approximations to the standard

Johnson-Cousins filters. The zero points on nights 1 and 2 were, within their errors, the same. The data for both nights were combined and new zero points determined. Secondary standards were chosen on single program frames from nights 1 and 2 and had their aperture magnitudes measured with DAOGROW in the same manner as the standards. As with the crowded standard frames, the secondary standard frames were reduced so as to leave in only the standards. After correction for exposure and airmass, the standard magnitudes of the secondaries were determined and the offset between PSF and standard magnitudes applied to the PSF magnitudes of all the stars on the frame. We were unable to calibrate the Field 5 $V$ frame as this was observed only on night 4 (See Tables 1 and 3). In lieu of calibration, we used the $I$, $(V-I)$ CMDs to choose an offset (by eye) for the Field 5 $V$ magnitude such that the color of its AGB/GB branch coincided with the AGB/GB branches of the other fields. This *ad hoc* calibration must be considered insecure.

No standards were observed through the $CN$ or $TiO$ filters, instead an offset was added to the instrumental $(CN-TiO)$ color so the mean color of stars bluer than $(V-I)=1$ was zero. The reasoning behind this is that we expect hot stars to have featureless spectra in the $CN$ and $TiO$ filter bandpasses. Due to the paucity of hot stars in Fields 1, 4 and 5, the mean $(CN-TiO)$ value is not well defined, and subsequently the $(CN-TiO)$ "calibration" of these fields is less certain.

## 3. COLOR-MAGNITUDE DIAGRAMS

In Fig. 2 we present calibrated $I$, $(V-I)$ CMDs for the five fields observed in M31. The number of stars in the plot, those matched between the $V$ and $I$ frames and with average (ALLSTAR) $\chi^2$ values (returned by DAOMASTER) less than 4, is indicated in each panel.

The bottom right panel of Fig. 2 shows the expected position of some features in the M31 CMDs. These features were placed using an absolute distance modulus of $(m-M)_0 = 24.41$ (Freedman & Madore 1990 (henceforth FM90), see also Secs. 3.2 and 4.2) and reddening and absorption values appropriate for Fields 1 to 4 (Sec. 3.1). The features shown are: (1) Fiducial sequences for a metal-poor (M15) and a metal-rich (47 Tuc) globular cluster (DaCosta & Armandroff 1990); (2) The Cepheid instability strip (RCP84, Kraft 1963); (3) Supergiants with spectral types M0Ib, M2Ib, M3Ib and M4Ib (RCP84, Lee 1970); and (4) Giants with spectral types M0II, M2II, M3II and M4II (RCP84, Lee 1970). We plot the globular cluster fiducials to give an idea of the luminosity of the giant branch tip, while the Cepheid instability strip shows that there are many potential Cepheids in our fields, especially Field 3.

Comparing the supergiant sequence to the CMDs, it appears that Field 3 has an excess of stars in this region compared to the other fields. An estimate of this excess can be made by counting stars brighter than an $I$ magnitude of 19 in our fields. In our Fields 1 to 5, we find 115, 175, 352, 123 and 113 stars brighter than $I = 19$. These counts were made using the $I$ data only, with no matching or $\chi^2$ criteria. The background field of RCP84 had 21 stars brighter than $I = 19$, which scales to $172 \pm 40$ stars for our fields when the relative field sizes are taken into account. The error in this result is the Poisson error in 21 (the number of stars in the background field of RCP84) scaled up appropriately by the ratio of field sizes. From these numbers, we estimate that Field 3 contains approximately 200 supergiants.

The most striking feature of Fig. 2 is the strong blue component seen in Field 3. This comes as no surprise as this field lies in a spiral arm and is currently undergoing star formation, as evidenced by the presence of Cepheids. Field 2 also has a blue component, though at a somewhat reduced level. We discuss these features more in Sec. 4.3.4 where we compare a census of AGB stars with star forming histories.

### 3.1. Reddening

Knowledge of the reddening is important as it affects many of the measurements made in this paper. For example, we count stars redder (intrinsically) than a certain color in our fields, which leads to the counts depending on the adopted field reddening. The reddening (via absorption) also affects the mean magnitudes of LFs, and consequently affects distance estimates and comparisons between LFs of the different fields.

The stars we observe in M31 suffer from both internal and foreground reddening. The scale on Fig. 1 shows the physical size of our fields in M31 to be ∼1.6kpc×1.6kpc. Within such large fields, the internal reddening will vary due to irregularities in the dust distribution. For example, an inspection of the Field 3 location in Hodge (1981) clearly shows dust lanes superimposed on the background. Even a field with a uniform dust distribution may have variable reddening due to the "depth" distribution of stars in the dust. This means that there is no single reddening value for a field, and that the reddening values we adopt are really average values.

With our data we can only make crude estimates of the reddening, and so we turn to previous studies to guide us. Hodge & Lee (1988), using $UBV$ photometry, investigated reddening as a function of galactocentric distance (in the plane) in M31. An inspection of their Fig. 17 suggests that the reddening in Fields 1, 2, and 4 (located at 4.6, 7.2, and 16.8 kpc) is around $E_{B-V} \sim 0.23$, while the reddening in Fields 3 (located at 10.8 kpc) is higher ($E_{B-V} \sim 0.4$) and in Field 5 (located at 31.5 kpc) lower ($E_{B-V} \sim 0.1$). We believe that the high reddening value in Field 3 indicated by Hodge & Lee's (1988) data might be inappropriate. The evidence in favor of this is: (1) The blue edges of the CMDs of Fields 2 and 3 both lie at similar colors (Fig. 2), indicating that the reddening in Field 3 is similar to that in Field 2; (2) In their study of the open star cluster C107 (Hodge 1979), which lies in Field 3, Bohlin *et al.* (1988) estimated the reddening of the cluster as $E_{B-V} = 0.20$; (3) A $BVRI$ investigation of Cepheids by Freedman and Madore (1990) indicated that the visual absorption in Baade's Field I (BF I) (in which Field 1 mostly lies) and BF III (within which Field 3 entirely lies) was 0.60 and 0.75 respectively. Adopting a value of 3.1 for the ratio of $A_V/E_{B-V}$ leads to reddening values of $E_{B-V} = 0.19$ and $E_{B-V} = 0.24$, suggesting a value for the reddening in Field 3 similar to the Field 1 value; and (4) In Sec. 3.2 we show good agreement between the position of the Cepheids and the instability strip in the $I$, $(V-I)$ CMD when a lower value of the reddening is adopted. Taking the above into consideration, we adopt a reddening of $E_{B-V} = 0.23$ for Fields 1, 2, 3, and 4. Beyond 10 kpc, Fig. 17 of Hodge & Lee (1988) shows that the reddening drops, the last point on this Figure indicates a reddening of $E_{B-V} = 0.13$ at a distance of approximately 20 kpc. The minimum reddening that Field 5 can possibly have is the foreground reddening of $E_{B-V} = 0.08$ (Burstein & Heiles 1984). As a compromise between the minimum reddening value and the reddening value at 20 kpc, we adopt a reddening value of $E_{B-V} = 0.10$ in Field 5.

We obtain $E_{V-I}$ from $E_{B-V}$ using the relationship, $E_{V-I} = 1.25 E_{B-V}$, given by Bessell & Brett (1988). The absorption in $V$ is derived by adopting a value of 3.1 for the ratio of $A_V/E_{B-V}$, and the absorption in $I$ (Cousins) is obtained from $A_I = 0.6 A_V$ (Cohen *et al.* 1981).

Reddening is unimportant in $(CN-TiO)$ for two reasons: (1) due to the small (300Å) separation between the central wavelengths of the $CN$ and $TiO$ filters, low levels of reddening have little effect on the $(CN-TiO)$ color; (2) the "calibration" of $(CN-TiO)$ ($<(CN-TiO)>= 0$ for hot stars) corrects for any reddening.

### 3.2. Cepheid Distance Modulus

Three of the M31 fields have regions in common with fields observed by Baade in which variable stars were identified. Variable stars in BF II are reported by Gaposchkin (1962), while Variable stars in BF I and BF III were identified by Baade & Swope (1965). Field 1 is mostly contained in BF I, while Fields 2 and 3 are totally contained in BF II and BF III respectively. Our outer two fields do not coincide with any of Baade's fields. From an inspection of Fig. 2 it appears that Field 4 may contain a handful of Cepheids, while Field 5 probably contains no Cepheids.

By locating the position of Field 1 on Plate I of Baade & Swope (1965) we found that Field 1 contains ten variable stars, five of which are Cepheids. Of these five Cepheids, Baade & Swope's identification of three was uncertain. The variable stars in Field 2 were identified by locating the position of Field 2 in Fig. 1 of Gaposchkin (1962). We counted 38 variables within the bounds of Field 2, and of these 15 are Cepheids.

The Plate and Figure we used to count the variables in Fields 1 and 2 were of insufficient detail to allow us to locate the variable stars on our frames. This was not the case for Field 3 though, on which we were able to find most of the variable stars using the more detailed finding charts (Plates IV and V) given by Baade & Swope (1965). In Field 3 we reidentified and obtained photometry of 80 variable stars, 60 of which are Cepheids. These 80 variable stars can be located using the Field 3 image, and Table 2, both of which are on the AAS CD-ROM series, volume XX, 1995. Using our photometric measurements, we plot in Fig. 3 a $I$, $(V-I)$ CMD of the identified Field 3 Cepheids, along with the expected position of the features described in Sec. 3.

Using the Cousins $I$-band Cepheid period-luminosity (P-L) relationship given by Madore & Freedman (1991, henceforth MF91),

$$M_I = -3.06(\pm 0.07)(\log P - 1.00) - 4.87(\pm 0.03)\ [\pm 0.18], \tag{1}$$

along with the periods from Baade & Swope (1965) and $I$ magnitudes from this study, we calculate an absolute distance modulus of $24.38 \pm 0.05$ for the Cepheids in Field 3. The P-L data for the Cepheids, along with the fit, are shown in Fig. 4. The fit was achieved by rejecting those stars, shown as open squares, which had deviations from the initial fit greater than twice the standard deviation of the deviations of the points away from the initial fit. The $I$ magnitudes shown have been corrected for 0.43 magnitudes of absorption in $I$. Our derived distance modulus to M31 is consistent with the value of $24.41 \pm 0.09$ derived from Cepheids by FM90 in BF III.

### 3.3. Giant Branch Width

The observed $(V-I)$ width of the giant branches (GBs) in Fig. 2 is a combination of four factors. Two of these, the age and abundance spread of the stars in the field, are "intrinsic" properties of the stars themselves. The other two factors, differential reddening and photometric errors, are "extrinsic" to the stars. If we can correct the observed GB width for contributions from the extrinsic parameters, we can use the intrinsic GB width to investigate (albeit in a crude manner) the star formation history of the field.

In the CMDs of Fig. 2 the GB blends in with the AGB. To disentangle the GB from the AGB we take advantage of the longer evolutionary timescale of the GB that causes stars to "pile-up" on the GB. This causes an "edge" in the CMD at the GB-AGB transition magnitude which, following Lee, Freedman & Madore (1993), we find by use of the Zero-Sum Sobel Kernel (ZSSK) edge detection algorithm. After rejection of stars which were clearly not on the GB or AGB we employed the ZSSK to determine the $I$ magnitude of the GB tip in our fields. A fuller description of the ZSSK can be found in Lee *et al.* (1993). The robustness of the ZSSK on our data was tested by using the ZSSK with bin sizes of 0.20, 0.10 and 0.05 mags, the results of which are summarized in Table 4. In Table 4 columns 2, 3 and 4 are the measured GB tip magnitudes (maximum value of the convolution of the field's stellar magnitude histogram (in $I$) with the ZSSK) for bin sizes of 0.20, 0.10 and 0.05 mags respectively, while column 5 is the average of these values. The resolution of the technique is determined by the bin width, though the technique will fail if the bins are so narrow that small number statistics dominate. Assuming the bin width to be an error estimator, we see from Table 4 that in Fields 1 and 2 we obtain consistent values. The measured GB tip magnitudes for Fields 3, 4, and 5 disagree between the 0.10 and 0.05 bins. When we use the 0.05 bins we find that in Field 5 there are 30 stars in the bins close to the maximum value of the convolution and for Field 4 this number increases to 200. At this level small number statistics will play a role in determining the tip, especially if the edge is ill-defined, as would be caused by an age or abundance spread in the stars on the GB. As the exact value adopted for the GB tip has little effect on the width measured for the GB, we simply adopt the average values given in column 5 of Table 4. Lee *et al.* (1993) employ the tip of the GB as a distance indicator, but warn that their calibration applies only to the metallicity range $-2.2 < [Fe/H] < -0.7$ and is unlikely to be applicable to a metal-rich population such as that found in M31's disk. In Fig. 2 the data is not deep enough to reach the main-sequence turn-off. This being the case, we note that the GB reaches from the magnitude limit of the data to the GB tip, as defined by the ZSSK. Knowing the location of the GB, we can now consider the factors contributing to its width.

An idea of one of the extrinsic factors which influences the width of the observed GB, differential reddening, can be arrived at by examining Hodge's (1981) *Atlas of the Andromeda Galaxy*. In Fields 4 and 5 there is no evidence of differential reddening: the background shows no patchiness and no dust clouds are apparent. In contrast, Fields 1, 2 and 3 appear patchy and have dust clouds outlined. We proceed by assuming that Fields 4 and 5 have no differential reddening, while the inner three fields do. Without knowledge of the level of differential reddening in the inner three fields, any attempt to interpret the GB width in terms of the intrinsic parameters will be necessarily limited.

To measure the contribution to the GB width from the other extrinsic parameter,

photometric error, we conducted ADDSTAR tests on our $V$ and $I$ frames. We first defined a "dispersionless" GB by rejecting stars (by eye) on the CMDs which obviously were not GB/AGB stars, and making a least-squares fit to the remaining stars. Artificial stars were generated along the dispersionless GB which were then broken down into $V$ and $I$ magnitudes. After generating random coordinates for these stars, they were added into the program frames using ADDSTAR. The ADDSTAR frames were reduced using a double pass (Sec. 2.2) and the two resulting ALLSTAR files were matched with the two ADDSTAR files using DAOMASTER, with the condition that a star had to be found in all four files to count as a recovery. The GB width of the recovered stars gives a measure of the photometric errors. We note that a similar method is often used to obtain photometric error estimates which can be compared to the analytical errors returned by ALLSTAR (e.g. Stetson & Harris 1988).

Results from the GB ADDSTAR tests in Field 4 are shown in Fig. 5. The leftmost panel in Fig. 5 shows the dispersionless relationship along which the stars were added, while the middle panel shows the added stars that were recovered and matched on both the $V$ and $I$ frames. The rightmost panel of Fig. 5 shows the real GB, as recovered from the frame on which the ADDSTAR tests were performed. A measure of the standard deviation, $\sigma$, of the color distribution of stars on the GB is given above each panel. To measure $\sigma$, we first dropped from consideration any stars brighter than the measured GB tip (column 5, Table 4) or fainter than an $I$ magnitude of 22.0 (the approximate limit of our data). The stars satisfying these magnitude criteria were split into ten bins in $I$ magnitude (binning in magnitude is necessary as the mean color changes with magnitude) and the mean and $\sigma$ of the color distribution were calculated for each bin. Outliers and non-AGB stars were rejected by excluding those stars that lay more than $2\sigma$ from the mean color, and then $\sigma$ was recalculated. To ensure that our rejection scheme was reasonable, the color bounds ($2\sigma$, all stars) for each bin were plotted for all the fields. These are shown in Fig. 5 for Field 4, where it can be seen that only outliers are being rejected. The $\sigma$ values shown above each panel in Fig. 5 are simply the average of the $\sigma$ measured in the ten magnitude bins. Measurements of the observed and photometric $\sigma$ measured in our five fields using the above techniques are given in columns 2 and 3 of Table 5. It is of interest to note that the value of $\sigma_{phot}$ (Column 3, Table 5) decreases as galactocentric distance in M31 increases. This result comes as no surprise, as we expect photometric errors to be greater in the more crowded inner fields.

If we assume that the observed distribution is a convolution of the intrinsic (metallicity and age spread) distribution and the distribution due to photometric errors, we can calculate the $\sigma$ of the intrinsic distribution according to $\sigma^2_{obs} = \sigma^2_{int} + \sigma^2_{phot}$. This is obviously not true for Fields 1, 2 and 3 where differential reddening makes a contribution to the width of the observed distribution. We include in Table 5 the calculated $\sigma$ of the intrinsic GB distribution (column 4) as well as the FWHM that corresponds to this $\sigma$ assuming a gaussian distribution (column 5).

In calculating the intrinsic FWHM values the effect of differential reddening in the inner-three fields were ignored. The level of differential reddening needed to account for the observed GB widths is large. For example, in Field 3 we measure the intrinsic FWHM of stars on the GB as $\sim 1$ magnitude in $(V-I)$, while the reddening, in $(V-I)$, is estimated at around 0.3 (Sec. 3.1). This suggests that differential reddening may not be the dominant factor contributing to the GB widths seen in these fields. As we have no way of dealing with the effects of differential reddening in the inner three fields we drop it from consideration and proceed by interpreting the intrinsic GB widths in terms of metallicity and mass variations.

What is the metallicity dispersion in our fields? We note that the metal-poor spheroidal component in M31 dominates the light (in $B$) for distances less than 4 kpc along the major axis (de Vaucouleurs 1958) meaning that only in Field 1 does it make a significant contribution. The other fields are dominated by the disk, and for these fields we adopt a dispersion of 0.165 dex. This is the value for the metallicity dispersion in the solar neighborhood, based on observations of 38 clusters in the solar vicinity by Cameron (1987). To see what effect the metallicity dispersion would have on the GB width we plotted the Z=0.05, 0.02, and 0.008 isochrones of Bertelli *et al.* (1994) for ages between 1 and 5 Gyr. In Fig. 6a we show these three isochrones for an age of $\sim$5 Gyr. The range in Z is equivalent to a range of 0.8 dex in metallicity, and we find that, regardless of age, the $(V-I)$ color difference between the Z=0.05 and the Z=0.008 isochrones is approximately 0.4. From this we conclude that a large variation in metallicity in Field 1 (presumably arising from the superposition of spheroidal and disk stars) could account for the width of its GB. In the outer four fields, we expect the GB width due to metallicity variations to be around 0.08. This is considerably smaller than the observed FWHMs of the GBs in the outer 4 fields and so we now turn to age for an explanation of the observed widths.

In a field which has had continuous star formation, the stars currently ascending the GB will be both younger, more-massive, and older, less-massive, stars. Regardless of metallicity, the loci of the GBs for the lower- and higher-mass stars will be different, causing a dispersion of the GB in the field. In Field 3, and to a lesser extent Field 2, the intrinsic width of the GB is greater than in the other fields. Both of these fields have strong blue components indicative of ongoing star formation. The obvious interpretation for the observed GB widths in these fields is that stars of different masses are "funnelling" onto the GB, their combined loci giving rise to the GB width. To estimate the mass range needed to account for the intrinsic GB widths in terms of mass funnelling, we use the solar metallicity (Z=0.02) isochrones of Bertelli *et al.* (1994). We find that for stars in the age range $1 \times 10^{10}$ yrs to $3.2 \times 10^{8}$ yrs (corresponding to turn-off masses of approximately 1 to $3\mathcal{M}_{\odot}$) the locus of the GB moves by around 0.6 in $(V-I)$. We illustrate this in Fig. 6b where we have plotted solar metallicity isochrones in the mentioned age range. Comparing this value to the FWHM values given in Table 5 it can be seen that, if this is the dominant source of GB width, Field 2 is dominated by stars in the mass range 1 to $3\mathcal{M}_{\odot}$ while Field 3, with its vigorous star formation, has a significant component of stars on the GB with masses greater than $3\mathcal{M}_{\odot}$. Fields 4 and 5, with their narrower GBs, probably ceased star formation some time ago and now their GBs are dominated by lower-mass stars than those of Fields 2 and 3.

We stress again that these results must be used as a guide only. A more detailed analysis would include other factors such as differential reddening, GB morphology and completeness. For example, with continuous star formation we would expect the GB color distribution to have an extended tail to the blue, the details of which would depend on the initial-mass function (IMF) in the field.

### 3.4. AGB Luminosity Functions

The AGB luminosity function (LF) is of interest as it provides insight into the formation and evolution of stars. For example, observations of the LMC showed a lack of bright AGB stars when compared to theoretical predictions (Iben 1981 and references therein). This discrepancy, known as the bright C-star mystery, led theorists to adopt revised mass-loss laws in their models for the more massive AGB stars, or to include effects such as semiconvection and convective overshooting. AGB evolution theory (Iben & Truran 1978, Renzini & Voli 1981) predicts that the tip luminosity of the AGB is a monotonic increasing function of mass, a result verified by Aaronson & Mould (1982). As pointed out by Reid & Mould (1984) this means that any point on the AGB LF will have contributions only from stars younger than a certain age. In summary, observations of the AGB LFs have the potential to furnish us with information on the star formation rate, the initial mass function and the mass-loss rate.

Assuming similar initial mass functions in the five M31 fields, a comparison of their AGB LFs allows us to compare the different star formation histories in these fields. For example, in a field that had ceased star formation a long time ago, we would expect a lack of brighter AGB stars, and would thus expect a steep LF. On the other hand, a field in which there has been continuous star formation would have luminous AGB stars, Cepheids and supergiants.

Calculating a single AGB star model is computationally expensive, while calculating an AGB LF for an ensemble of stars is unrealistic. Much of the computational expense involved in calculating an AGB LF can be circumvented through the use of empirical relationships, a method known as synthetic AGB evolution. The AGB LFs produced by such models are in terms of bolometric magnitude, and observers often derive bolometric AGB LFs to allow comparison with synthetic models. To allow easy comparison between synthetic models and previous studies, we also derive bolometric AGB LFs. To obtain the bolometric magnitude of a (late-type) M31 star, a bolometric correction (BC) is applied to the $I$ magnitude. Adopting $(m-M)_0 = 24.41$ (FM90, see also Secs. 3.2 and 4.2), and applying the BC given by Bessell & Wood (1984) we have

$$M_{\rm bol} = I_0 + 0.3 + 0.38(V-I)_0 - 0.14{(V-I)_0}^2 - 24.41. \tag{2}$$

This BC is slightly inappropriate for the C-stars, though we use it on all the stars to avoid any spectroscopic bias.

The completeness in the $M_{\rm bol}$, $(V-I)_0$ plane was determined by Monte-Carlo methods. We follow Reid & Mould (1984) who, by consideration of the M92 giant branch tip, defined an AGB star as having $(V-I)_0 > 1.48$. Stars were added in the $M_{\rm bol}$, $(V-I)_0$ plane in the region defined by $(1.48 < (V-I)_0 < 3.98)$ and $(-6 < M_{\rm bol} < -1)$. The stars are mapped to unique $V$ and $I$ instrumental magnitudes, and added into the $V$ and $I$ frames. The frames were then reduced using a double pass as described in Sec. 2.2. The files needed to determine completeness corrections can be obtained by matching the ADDSTAR and ALLSTAR files, and filtering the resulting match file appropriately. The region to which stars were added in the $M_{\rm bol}$, $(V-I)_0$ plane was split into a 10×10 grid and the completeness determined for each element of the grid. Fields 5 and 4 had 4000 stars added to them over 4 tests, while Fields 3, 2 and 1 had 4000 stars added over 2 tests. Adding stars at this level did not significantly increase the crowding.

To correct the LFs for stars along the line of sight, stars from the background field of NGC 205 (RCP84, Fig. 6) were mapped into the $M_{\rm bol}$, $(V-I)_0$ plane. No mention of the completeness of the NGC 205 background CMD is made by RCP84, though their $I$ magnitude limit appears to be around 21. As we are mainly interested in stars brighter than $M_{\rm bol} = -3.5$ ($I_0 \sim 20.5$), we proceed by assuming their CMD to be complete. The background counts are scaled to allow for our field size and subtracted from the completeness corrected counts. Fig. 7 shows the corrected counts in Field 4 after background subtraction. Some of the elements in Fig. 7 show negative counts, this is simply due to scaling the background counts, and is compensated for when the counts are summed in $M_{\rm bol}$ over the color bins.

In Fig. 8 we show the bolometric LFs for the five M31 fields before (triangles, broken lines) and after (squares, solid line) corrections for completeness and background counts. The errors shown for the uncorrected LFs are $\sqrt{n}$ errors, while the errors for the corrected LF take into account the errors associated with the incompleteness and background corrections. Bins were only plotted if more than 10 real stars (no background corrections) were in the bin, and are open squares if the completeness was less than 50%. The turnover in the LFs is expected as at faint magnitudes the stars are too blue to meet the aforementioned color criterion. The completeness of the data will also play a role in determining the turnover. The Field 5 LF has very few stars and is strongly affected by corrections for background counts, as can be clearly seen. We note that in all LFs, except Field 2, the bins with $M_{\rm bol} < -3.5$ have a completeness level greater than 50%, and that completeness corrections are unimportant for the AGB stars.

We can now ask how the AGB LFs differ in the fields. In the following comparison we drop Field 5 from consideration due to its lack of stars and ill-defined LF. The corrected LFs in Fig. 8 for Fields 1 through 4 were normalised such that they all contained the same number of stars in the magnitude range $-5 < M_{\rm bol} < -3$. In Fig. 9 these LFs are plotted in the range $-6 < M_{\rm bol} < -3$. We exclude errors from this plot to avoid crowding though errors are included in Fig. 8. The most obvious feature in Fig. 9 is that Field 3 has the shallowest AGB LF gradient. Field 3 has a strong blue component and contains many Cepheids and supergiants (see Fig. 2 and Sec. 3) indicating that it is currently undergoing star formation. The shallower AGB LF gradient suggests that the increased level of star formation has been going on in Field 3 for *at least a few Gyr.* Field 4 has the next shallowest gradient, though has a reduced blue component compared with Fields 2 and 3 (again, see Fig. 2). This suggests that Field 4 may have had more active star formation in the past. The AGB LFs of Fields 1 and 2 look similar, even though Field 2 has a moderate blue component whereas Field 1 has virtually no blue stars. This is suggestive that the star formation that we see in Field 2 has not been ongoing. To understand the AGB LFs more fully, synthetic AGB models need to be generated, which is beyond the scope of this paper. In the study of M31 by RCP90 a similar trend was seen, with a field at $R_{\rm M31} = 4$ kpc having an older component than fields at 11 and 20 kpc.

In Sec. 3.3 we suggested, from an investigation of GB widths, that differences existed between the mass ranges of stars on the GBs in the five fields. We showed that after correcting for photometric errors, and ignoring the possibility of differential reddening, that Field 3 had the widest GB. Using the isochrones of Bertelli *et al.* (1994) we then demonstrated that the width could be due to a large range in mass of stars on the GB. This result is in agreement with the shallower AGB LF observed in Field 3. Given this result, it is unexpected that Field 2, with a GB wider than all but the Field 3 GB, does not have the next shallowest AGB LF. It may be that in this field the AGB LF is dominated by lower-mass stars, and the brighter AGB stars provide insufficient "leverage" to give the field a shallower LF. Field 4 has a GB width narrower than that of Fields 2 and 3, though its AGB LF has a gradient shallower than Field 2, and steeper than Field 3. This result would seem in accord with the previous suggestion that Field 4 has a large intermediate aged population. As the GB width of Field 1 is hard to interpret because of the possibility of large metallicity effects, and the AGB LF of Field 5 suffers from small number statistics we present no comparison of AGB LFs and GB widths in these fields.

In conclusion, *we see different AGB LFs in the different fields, indicative of different star formation histories.* The AGB LFs will be further discussed in Sec. 4.3.4 where we compare AGB LFs and a census of AGB spectral types.

## 4. THE COLOR-COLOR DIAGRAM: AN AGB CENSUS

If a star is sufficiently massive, as it evolves up the AGB it can undergo thermal pulses which can lead to third dredge-up. This can cause it to change the chemistry of its outer envelope from oxygen-dominated (M-star) to carbon-dominated (C-star). The intermediate class between these are S-stars wherein the abundances of oxygen and carbon are similar.

To find the C-, M-, and S-stars in our five fields, the photometric data is used to plot $(CN-TiO,\ V-I)$ color-color diagrams. Carbon stars, having strong CN bands and weak TiO bands, have large positive values of $(CN-TiO)$. Conversely, M-stars have strong TiO bands and weak CN bands, at least compared with C-stars, and this gives them large negative values of $(CN-TiO)$. When $(CN-TiO)$ is plotted against $(V-I)$, a temperature discriminator, a clear bifurcation is seen between C- and M-stars. This can be seen in Fig. 10 in which we plot color-color diagrams for all five fields. At colors redder than $(V-I)_0 = 2$ the bifurcation in $(CN-TiO)$ is as great as a magnitude, making the distinction very clear. S-stars have C $\sim$ O and are expected to lie somewhere between the C- and M-stars in the $(CN-TiO,\ V-I)$ color-color diagram.

### 4.1. Selection Criteria

How well does this system work? Follow up spectroscopy of candidates in Fields 3 and 4 by Brewer, Crabtree & Richer (1995) has shown the technique to be highly successful at identifying C-, M- and S-stars: of 17 candidate C-stars in Field 3 redder than $(V-I)_0 = 1.8$, 15 were spectroscopically confirmed as C-stars. The remaining two were difficult to classify due to weak spectra. Spectral observations (Brewer, Crabtree & Richer 1995) show that stars for which $(V-I)_0 > 1.8$ and $(CN-TiO) > 0.3$ are C-stars. We define M-stars as being those with $(V-I)_0 > 1.8$ and $(CN-TiO) < -0.2$, which we justify by noting that a star of spectral type M3 in Field 3 lies at $(V-I)_0 = 1.83$, $(CN-TiO) = -0.22$. These M-star and C-star color limits are shown as solid lines in the lower right panel of Fig. 10 (The dashed lines on Fig. 10 will be discussed in Sec. 4.3.2). Though our definitions of what constitutes C- and M-stars differ slightly from other studies, we point out that we are internally consistent.

The M-star region on the color-color diagram can suffer contamination from both M31 M-supergiants and Galactic M-stars. The brighter of these stars can be eliminated by applying a magnitude criterion to the color-color diagram. In their study of M31, RCP90 found that the number of stars brighter than $I = 19.0$ in their program field was similar to the number in the background field of RCP84. This suggests that stars in our fields with $I < 19.0$ are either foreground stars or M31 supergiants. In Fig. 10 we applied a similar criterion and excluded those stars with $I_0 < 18.5$. The exclusion of these brighter stars does not preclude the possibility of contamination from Galactic M-dwarfs. At a Galactic latitude of $b = -21^\circ\!.6$, M31 suffers from modest foreground contamination. The number of stars in the background field CMD from RCP84 redder than $(V-I) = 1.8$ and fainter than $I = 19.0$ is 12, which when scaled to allow for the relative field sizes in RCP84 and the present work, becomes $100 \pm 30$. Counting stars redder than $(V-I) = 1.8$ and fainter than $I = 19.0$ in the CMDs we find 19371, 3311, 4214, 8086, and 372 stars in Fields 1 to 5 respectively. It is clearly seen that M31's stars dominate the sample in all but Field 5 where we estimate that approximately 25% of the stars with the aforementioned color and magnitude criteria are field stars.

Contamination in the C-star region of the color-color diagram from dwarf C-stars in the Galactic halo is of no concern as their low space density ($0.019^{+0.044}_{-0.016}$ deg$^{-2}$, Green 1992) renders them statistically unimportant. Any Galactic C-stars will be excluded by the magnitude criterion mentioned in the previous paragraph.

We are interested in the C- and M-stars found on the AGB. In their review article, Iben & Renzini (1983, henceforth IR83) state that for a star to develop its own peculiar abundances it has to be brighter than $M_{\rm bol} \sim -3.5$. Using the BC from Bessell & Wood (1984), given in Eq. (2), we calculated the bolometric magnitude of the M-stars, and excluded those fainter than $M_{\rm bol} = -3.5$. As pointed out by Bessell & Wood (1984) though, their BC is not applicable to C-stars. The magnitude criterion we adopted for C-stars was $I_0 < 20.6$. This was chosen after noting that the Field 3 and Field 4 color-color diagrams had well discriminated C-stars, and none of these were fainter than $I_0 = 20.6$. An inspection of Fig. 10 suggests that many of the Field 1 C-star candidates are interlopers, as there is a blueward skew in the color distribution compared to the other fields and many of the candidate C-stars in Field 1 fail the magnitude criterion (see Table 6). Though the C-star magnitude criterion is not entirely effective at eliminating blue interlopers, it does eliminate the more obvious ones.

The candidate C-stars in each field (stars with $(V-I)_0 > 1.8$, $(CN-TiO) > 0.3$, and $18.5 < I_0 < 20.6$) are listed in Table 7 which is presented in its complete form on the AAS CD-ROM Series, volume XX, 1995. The following data appear for each C-star candidate: in column (1) an ID for each star, ordered by magnitude; in column (2) X-Y coordinates (these can be used to identify the candidates on the $I$-band images, included on the same volume of the AAS CD-ROM); in column (3) the $I$ magnitude along with its ALLSTAR error; in columns (4) and (5) the $(V-I)$ and $(CN-TiO)$ colors, along with their errors derived by adding ALLSTAR errors in quadrature. In Table 8 (again, presented in its complete form on the AAS CD-ROM Series, volume XX, 1995) we list the M-stars (stars with $(V-I)_0 > 1.8$, $(CN-TiO) < -0.2$, $I_0 > 18.5$, and $M_{\rm bol} < -3.5$). The layout of Table 8 is the same as Table 7.

In Fig. 11 we plot the CMDs of those stars with $(V-I)_0 > 1.8$, $(CN-TiO) > 0.3$, and $18.5 < I_0 < 20.6$; the candidate C-stars. A comparison of Figs. 2 and 11 shows the C-stars to be among the brightest and reddest stars in the fields. Also included in Fig. 11 are the Cepheids identified in Field 3. A comparison of the C-stars and Cepheids in Field 3 shows that they have similar $I$ magnitudes, suggesting that C-stars are potentially useful distance indicators.

### 4.2. C-Stars: Smoky Standard Candles?

Due to their high luminosity, very red color and strong spectral characteristics, C-stars are potentially good distance indicators. One of their advantages over Cepheids, for example, is that they can be identified and used as a distance indicator in a single epoch of observation. Other advantages are the reduced effect of interstellar reddening when observing at longer wavelengths where C-stars are brightest and, unlike M-stars, there is only a remote chance of contamination from dwarf field stars. One possible caveat of using the C-star LFs as standard candles is the need to adopt a universal C-star LF which assumes a common IMF and star-formation rate (SFR). This aside, there is strong observational evidence that there is a universal C-star LF for certain types of systems. Richer, Pritchet and Crabtree (1985) concluded that if a galaxy is sufficiently metal-rich ([Fe/H] $> -1.8$) and sufficiently bright ($M_V < -12.9$), then its C-stars will have the same luminosity as those in Fornax, the Magellanic Clouds, the Milky Way, M31 and NGC 205.

The four-band technique used here is limited by seeing and crowding conditions; enough positive identifications of C-stars are needed to build a LF to allow a reasonable estimate of the mean to be obtained. For example, the innermost field in this study is of no use for determining the distance modulus of M31 as the crowding produces errors in the photometry which cause many misidentifications and a deviant C-star LF. Taking this effect into account, we estimate that the current technique may be workable out to the Virgo cluster galaxies. We back up this claim by noting that in Fig. 11 the C-stars are as bright as the Cepheids in $I$, and that Cepheids have recently been identified in the Virgo cluster with both ground-based (Pierce *et al.* 1994) and HST (Freedman *et al.* 1994) observations.

In earlier work by the Richer group, a mean absolute Cousins $I$ magnitude of –4.75 was adopted for C-stars. This value was derived by transforming Richer's (1981) photometry of 70 C-stars on the Johnson system in the LMC Bar-West (BW) field (originally identified in the survey of Blanco, Blanco & McCarthy 1978) to the Cousins photometric system using the transformation given by Bessell (1979). After allowing for reddening and absorption in their LMC BW field (Richer, Pritchet & Crabtree 1985, and references therein), and adopting a distance modulus for the LMC of 18.45 (Welch *et al.* 1986) a mean absolute Cousins $I$ magnitude of –4.75 ($\sigma = 0.40$) was derived. The transformation given by Bessell (1979), $(V-I_C)_0 = 0.835(V-I_J)_0 - 0.13$, is valid for stars with $2.0 < (V-I_J)_0 < 3.0$. The mean $(V-I_J)_0$ color of Richer's (1981) stars was 3.0 with a standard deviation of $\sigma = 0.6$ meaning that extrapolation of Bessell's (1979) transformation equation was used in many cases. To verify that the extrapolation is valid, we made a comparison with Cousins $I$ photometry of the same 70 stars made by Blanco, McCarthy & Blanco (1980). After correcting the mean Cousins $I$ magnitude of Blanco *et al.* (1980) for the previously mentioned absorption and distance modulus, a value for the mean absolute magnitude of the LMC BW C-star sample of –4.79 ($\sigma = 0.35$) is derived. The agreement between the values shows that the extrapolation of Bessell's (1979) transformation was valid. For consistency with previous papers, we adopt the value of –4.75 for the mean absolute magnitude of the LMC BW C-stars.

Figure 12 shows $I_0$ LFs for the C-stars from all five M31 fields and the LF for the LMC BW C-stars (with good photometry) from Richer (1981). The stars in the M31 LFs pass the color and magnitude criteria described in Sec. 4.1, and have been corrected for absorption as outlined in Sec. 3.1. The LMC BW C-star magnitudes were corrected for absorption and moved to the M31 distance modulus by adding the difference between the M31 and LMC absolute distance moduli (FM90, Welch *et al.* 1986). For the M31 C-stars in Fig. 12 we give $< I_0 >$, the number of stars, the standard deviation and the error in the mean for each distribution.

As previously discussed, it is apparent for Field 1 that there is a lot of contamination from stars scattering into the C-star region. As this is hard to disentangle, we drop Field 1 from further discussion. Field 5 we also drop from consideration due to the paucity of C-stars and, again, problems with non-C-stars scattering into the region. For a combined sample of the Field 2, 3 and 4 C-stars, we find a mean value of $19.61 \pm 0.03$ (error in the mean). Taking the mean absolute $I$ magnitude of C-stars to be –4.75, as previously discussed, gives an absolute distance modulus of $24.36 \pm 0.03$ for M31, in agreement with the value of $(24.45 \pm 0.15)$ found by RCP90.

The distance modulus derived from the C-star LFs $(24.36 \pm 0.03)$ is in accord with that derived from the Field 3 Cepheids $(24.38 \pm 0.05)$ in Sec. 3.2. We note that the distance modulus derived from the Field 3 C-star LF agrees with the Cepheid distance modulus *regardless* of the reddening value (see Sec. 3.1) adopted in Field 3.

With the adopted reddening values in the fields, the mean magnitudes of the C-stars in Fields 2, 3 and 4 are, within the errors in the means, identical. The similarity of the C-star LF in arm (Field 3) and interarm (Fields 2 and 4) regions argues against a dependence of the C-star LF on IMF, SFR and metallicity. From these results we conclude that *the C-star LF provides a robust standard candle.* As discussed briefly in the introduction, this then implies that C-stars are useful as probes of Galactic structure, as it appears that neither star formation history or metallicity gradient in the disk of the Milky Way will affect the C-star LF.

Figure 12 also supports the hypothesis that higher mass stars avoid becoming C-stars (IR83). IR83 mention that Cepheids are the progenitors of bright ($M_{\rm bol} < -6$) AGB stars, and that the lifetimes of Cepheids and bright AGB C-stars are similar ($\sim 10^6$ yr). This leads us to expect comparable numbers of bright C-stars and Cepheids in our fields. In fact, what we see is that the C-star LFs of Fields 3 and 4 are alike despite Field 4 having fewer Cepheids than Field 3 (see Fig. 2). This argues that the lack of bright C-stars (the "bright C-star discrepancy") is not due to a lack of massive progenitors (the "old age solution") and that either the models are somehow incomplete or the bright C-stars are very red and are being missed by our survey. A strong argument against the second of these possibilities was provided by Frogel & Richer (1983) who showed a scareness of luminous AGB stars in the Bar West field of the LMC.

### 4.3. The C/M ratio: A Significant Statistic

Blanco, Blanco & McCarthy (1978) noted that the number ratio of C-stars to M-stars (henceforth C/M) was much greater in the metal-poor Magellanic Clouds than it was in the metal-rich Galactic Nuclear Bulge. They attributed the different C/M ratios to differences in metallicity or age, or to both of these parameters. Presently, the mechanism controlling the C/M ratio is believed to be the metallicity of the gas out of which the stars condensed, though it remains unclear what effect parent galaxy morphology, star formation history, and age play (Pritchet *et al.* 1987). In systems such as globular clusters and giant elliptical galaxies the C/M ratio is undefined as it is believed that stars with masses $\lesssim 0.9\mathcal{M}_{\odot}$ terminate their evolution before being able to undergo helium shell flashing (Renzini 1977).

### *4.3.1. The "Raw" C/M Ratio*

In Fig. 10 we presented color-color diagrams for the five fields in M31 in which it was apparent that the C/M ratio increased with $R_{\rm M31}$. To get a more quantitative picture of this increase we count the C- and M-stars in the five fields using the color and magnitude criteria described in Sec. 4.1. The resulting counts are presented in Table 6. The first two columns in Table 6 give field number and $R_{\rm M31}$, columns 3 and 4 give the counts when the C- and M-stars are selected by color criteria only, and columns 5 and 6 contain counts when the C- and M-stars are selected using the color and magnitude criteria. A comparison of columns 3 and 5 of Table 6 suggests that in Fields 1 and 5 we have a lot of scatter, while in Fields 2, 3 and 4 we have very little.

Before determining the C/M ratios we multiply the M-star counts in Field 5 by 0.75 (see Sec. 4.1) to correct for foreground stars. In Fig. 13 we plot the C/M ratios as a function of $R_{\rm M31}$. The error bars in Fig. 13 are propagated from the Poisson counting errors of the C- and M-star counts. We note that the C/M ratio for Field 1 is likely an upper limit as its C-star LF (Fig. 12) is skewed to fainter magnitudes suggesting that many of the C-star candidates in Field 1 may be spurious.

The correlation seen between C/M and $R_{\rm M31}$ in Fig. 13 will be discussed in Sec. 4.3.4 after we have considered what effects completeness may have on the data.

### *4.3.2. Effects of Completeness on the C/M Ratio.*

What are the potential effects of completeness on the measured C/M ratio? This question was briefly discussed by Cook, Aaronson and Norris (1986) who showed, using data from Blanco, McCarthy & Blanco (1980), that the C/M ratio measured in the LMC

Bar West and Radio Center fields changed with sample depth in $I$. In this section we consider the effects of completeness on our data in more detail.

Theory predicts that under certain conditions an M-star will evolve into an S-star and then a C-star as it ascends the AGB. As a star ascends the AGB it becomes cooler and more luminous. As we expect C-stars to be generally bolometrically brighter than M-stars it might be naively expected that the C-stars are more complete in our data. This is not necessarily the case. The reddest C-stars may suffer from incompleteness on the $V$ frame despite their high bolometric luminosity. Indeed, we would fail to detect bolometrically bright infrared C-stars at all. The $CN$ and $TiO$ filters further complicate the issue as these were designed to exploit the fact that the molecular opacities in C-stars makes them brighter in $TiO$ than $CN$ while the molecular opacities of M-stars makes them brighter in $CN$ than $TiO$; the difference between the $(CN-TiO)$ color for C- and M-stars can be as great as a magnitude (see Fig. 10). As a star has to be found on the $CN$, $TiO$, $V$ and $I$ frames to be identified, we have to consider the completeness on all four of these frames.

As a demonstration that completeness can have a significant effect we present in Fig. 14a the $I$-band LFs for the C- and M-stars in Field 3 that were selected according to the criteria given in Sec. 4.1. The bin size used to produce these LFs is 0.2 mags, and the M-star counts have been multiplied by 0.1, so as to fit on the same diagram as the C-star LF. Figure 14a shows that the C-star LF has a brighter mean magnitude than does the M-star LF. This result is predicted by the third dredge-up theory for the formation of AGB stars which shows that, under some circumstances, M-stars evolve into C-stars as they ascend the AGB (IR83).

As the C- and M-star LFs are different, completeness will affect the measured C/M ratio. This is shown in Fig. 14b where we plot the logarithm of the cumulative counts of C- and M-stars that are brighter than the abscissa magnitude. In Fig. 14b the logarithm of the C/M ratio at a limiting $I$ magnitude is the (negative) “gap” between the distributions. If the C- and M-stars had identical LFs, though different scaling factors, the lines in Fig. 14b would have a constant separation, this separation being dependent on the scaling factors of their LFs. For the Field 3 C- and M-stars shown in Fig. 14b the gap (and hence the C/M ratio) between the distributions varies as a function of limiting magnitude. The variation of C/M as a function of limiting magnitude is shown in Fig. 14c where it is seen (from right to left) that initially the C/M ratio increases as fainter stars are excluded. This increase is expected as the C-star LF in $I$ is brighter than the M-star LF, as shown in Fig. 14a. At even brighter magnitudes ($I_0 < 19.5$) the C/M ratio decreases. This is explained by the tail of the M-star distribution at brighter magnitudes.

In summary, we have shown that the measured C/M ratio from Field 3 is sensitive to the $I$ magnitude limit of the data. As completeness can potentially affect the measured C/M ratio, it needs be taken into consideration when counting the C- and M-stars.

Every star in the color-color diagrams of Fig. 10 has been found and fitted on the $CN$, $TiO$, $V$ and $I$ frames, and subsequently successfully matched. To measure the completeness of the stars in the color-color diagram, artificial stars are added to, and recovered from the $CN$, $TiO$, $V$ and $I$ frames. Using this Monte-Carlo method allows completeness to be measured without consideration of the many (often unknown) parameters needed for an analytical method to succeed. The Monte-Carlo method of adding artificial stars has been extensively used in the past to investigate monochromatic LFs (e.g. Brewer *et al.* 1993, and references therein) and bolometric LFs (Hudon *et al.* 1989, this paper). Unlike these previous studies, the present study suffers from a degeneracy of stars in the $(CN-TiO,\ V-I)$ plane, i.e. a point in the plane does not represent unique $CN$, $TiO$, $V$ and $I$ magnitudes. To obtain the magnitude of a star in the color-color plane a relationship between color and magnitude is needed. This can be provided by CMDs of the C- and M-stars. As this is the first time that $(CN-TiO,\ V-I)$ completeness tests have been done, we describe below our methodology.

In Sec. 4.1 the color criteria we used to define regions on the $(CN-TiO,\ V-I)$ plane where we would find C- and M-star candidates were unbounded inequalities in $(V-I)$ and $(CN-TiO)$. These criteria lead to there being two infinite areas on the $(CN-TiO,\ V-I)$ plane where stars are classified as C- or M-type. As it is impossible (and also unnecessary) to measure the completeness of the entire C- and M-star regions defined in this way, we use further color criteria to constrain the C- and M-star regions to finite sizes on the $(CN-TiO,\ V-I)$ plane.

To the criteria from Sec. 4.1 we add a red limit (for both C- and M-stars) from the reddest, $(V-I)_0 = 3.1$, C-star for which a spectrum was obtained (Brewer, Crabtree & Richer 1995). The lower $(CN-TiO)$ bound for C-stars was similarly defined by spectroscopic observations while the upper bound was arbitrarily set so as to encompass all C-stars apart from obvious outliers. The upper $(CN-TiO)$ bound for M-stars was defined using spectroscopy and the lower bound was set to $-0.7$, so as to have the same width in $(CN-TiO)$ as the C-star region. Although some M-stars lie out of this region, there are enough in the region for a statistically significant comparison, and it is the relative numbers from field to field that are important. The C- and M-star regions are shown in Fig. 10 as the areas bounded by the solid and dashed lines in the lower right panel. Changing the definition of what constitutes C- and M-stars, will cause the C/M ratios to have different values to those previously calculated.

Artificial C- and M-stars were generated by randomly placing points in the C- and M-star regions of the $(CN-TiO,\ V-I)$ plane. The $(V-I)$ and $(CN-TiO)$ colors of these artificial stars are broken down into representative magnitudes using the $I$, $(V-I)_0$ and $TiO$, $(CN-TiO)$ CMDs of the real C- and M-stars from these regions. The stars in the C-star region of the Field 1 color-color diagram were found to be fainter than the other candidate C-stars, so a sample representative of the Fields 3 and 4 C-stars was used. In Field 5 the paucity of C- and M-stars led us to adopt mean color-magnitude relationships from other fields. When the artificial colors are broken down into magnitudes, we ensure that the dispersion of the artificial data matches that of the real data. Random X-Y coordinates are generated for the artificial stars, and the ADDSTAR routine in DAOPHOT II is used to scale PSFs to the appropriate magnitudes and add the stars into the $CN$, $TiO$, $V$ and $I$ frames.

The four frames are reduced using a double pass (Sec. 2.2) and the resulting four ALLSTAR files were matched (by position) with the four ADDSTAR files using DAOMASTER. By imposing the condition on DAOMASTER that a star had to be found in *at least* four of the eight files resulted in the rejection of real stars that were found on less than four frames. The file of matched stars from DAOMASTER was filtered to produce three files which contained the added stars, the added stars that were recovered, and the real stars.

For each field we ran two completeness tests, and on each we added 500 C-stars and 500 M-stars. Adding these stars to each frame resulted in approximately 1 star being added to every $60\times60$ pixel section of the frame. Visual inspection showed this to be a reasonable level; the added stars did not overlap with stars of a similar magnitude. Further confirmation that we were not drastically altering the crowding characteristics of the frames came from noting that the numbers of C- and M-stars recovered from the program frames without stars added in were similar to the numbers recovered from the frames with stars added in.

We now have all the synthetic data needed to determine completeness corrections. As the colors and magnitudes of stars have a dependence on position in the $(CN-TiO,\ V-I)$ plane, we bin the C- and M-star regions into 4 bins in $(V-I)$ and 3 in $(CN-TiO)$. For each of these 24 bins the completeness is calculated by counting the number of stars added to the bin, and the number of these that were recovered. A star added in one color bin may be recovered in a different bin, an effect known as "bin jumping" (Drukier *et al.* 1988). In matching our added and recovered stars we imposed no color criteria on the recovered stars. This is reasonable as we found the scatter around the bins to be approximately symmetrical, with as many stars jumping into a bin as jumping out. Having obtained the completeness, we next wish to apply the value to the number of real stars in the bin. The real stars in each bin are counted and then the completeness corrections applied. The real stars are from the ADDSTAR frames as the corrections apply *only* to the frame on which they were calculated. All of the added, recovered and real stars were only considered if they pass the color and magnitude criteria (Sec. 4.1) to be C- or M-stars.

The results of the incompleteness tests on Field 3 are summarized in Fig. 15. This Figure shows the number of real stars in each bin followed by our calculated completeness for typical stars found in the bin. The lower numbers in the bins are the corrected counts. The errors include both contributions from Poisson errors in the number of real stars and the error in the completeness correction (Bolte 1989). As can be seen in Fig. 15 the redder M-stars are more incomplete than the bluer M-stars. This result is contrary to what might be expected, since as an M-star evolves up the AGB it is generally expected that it will become redder and more luminous. There are two effects which could cause this unexpected result. The cooler M-stars should have stronger $TiO$ bands, and will subsequently be fainter in the $TiO$ frames, and possibly less complete. The other possibility is that these M-stars could also be less complete due to being fainter in $V$.

In Fig. 16 we plot C/M ratio against $R_{\rm M31}$ for stars that were were bounded by the color criteria and uncorrected (triangles) and corrected (squares) for completeness. We point out that the uncorrected C/M ratios in Figs. 16 and 13 are different as additional color criteria were used to define the C- and M-stars in Fig. 16. Figure 16 shows that, after completeness corrections, the C/M ratios in Fields 1 and 5 increased, while they decreased in Fields 2, 3 and 4. In Fields 1 and 5 we found that the C-stars were less complete than in the other fields. Though the cause of this is unknown, it may be due to reasons such as the $CN$ frames being shallower (poorer seeing, brighter sky) or problems with the $(CN-TiO)$ calibration in these fields.

In conclusion, we have shown that the completeness has an effect on the measured C/M ratios, but in the case of our data, the effect is small and the trend seen between C/M and $R_{\rm M31}$ persists. This shows that the increase of the C/M ratio with $R_{\rm M31}$ is a true physical effect, and not due to some obscure effect related to the crowding statistics in the fields. For the rest of the paper we consider the "raw" C/M ratios shown in Fig. 13. The reasons for this are threefold. Firstly, as previously mentioned, in Fields 1 and 5 we adopted a mean color-magnitude relationship for the C-stars which may be invalid. Secondly, some bona fide C- and M-star candidates are excluded from the "corrected" counts as they lay outside the closed C- and M-star regions. And finally, the conclusions arrived at using the "raw" ratios are similar if the "corrected" C/M ratios are used instead.

### *4.3.3. Comparison With Previous Studies*

The C/M ratio in M31 has been previously investigated by RC85, Cook, Aaronson & Norris (1986) and RCP90. RC85 investigated a disk field in BF II at $R_{\rm M31} = 11$ kpc. Using the four-band technique they identified 5 C-stars and 41 stars later than type M5 in a $1.2' \times 2'$ field. The ratio of 5/41 was, at the time, the smallest C/M ratio found in any extragalactic system. Interpreting this ratio as a metallicity effect led RC85 to conclude that the metallicity of their field was approximately solar. Cook *et al.* (1986) investigated two $1.5' \times 2.5'$ fields, one in BF II at $R_{\rm M31} = 7$ kpc and the other in BF III at $R_{\rm M31} = 10$ kpc. They found only one C-star in each field and concluded that they were measuring primarily an old disk population which was unaffected by the presence of extreme population I. RCP90 looked at two $2.2' \times 3.5'$ disk fields; one in BF IV at $R_{\rm M31} = 20$ kpc was observed through $CN$, $TiO$, $V$ and $I$ filters, while the other, at $R_{\rm M31} = 4$ kpc, was observed through $V$ and $I$ filters only. After correcting for field stars RCP90 obtained a C/M ratio for their field at 20 kpc of 6/39, which was identical, within the errors, to the ratio found in the field at 11 kpc by RC85. This result was unanticipated by RCP90; if the C/M ratio depended solely on metallicity a much higher ratio ($\sim$0.3) would have been expected in the outer field if the metallicity gradient of Blair, Kirshner, & Chevalier (1982) was extrapolated out to the field at 20 kpc. This result was attributed by RCP90 to either small number statistics, a flattening out of the metallicity gradient or to star formation differences. We note in passing that the total area covered by all previous four-band surveys in M31 was 17 $\Box'$, while the present study surveys an area of 245 $\Box'$.

### *4.3.4. The C/M Ratio: An Alternative Abundance Indicator?*

Figure 10 shows color-color diagrams for all the five fields. The panels for Fields 1 and 5 show that it is possible that the C-star candidates in these two fields may just be result of photometric errors. If indeed this is the case, and there are no C-stars in Fields 1 and 5, there are two possibilities. The lack of C-stars could be explained by a very metal-rich environment, or a cessation of star formation sufficiently long ago that stars now ascending the AGB do not undergo thermal pulsing. The latter possibility seems unlikely in Field 1 as we know that it does contain bright AGB stars (see Fig. 8). Field 5 also contains bright AGB stars (Fig. 8) though, after correction for foreground stars, at low levels. To rule out the possibility that in Field 5 we have an old population we need spectra to confirm the C-star candidates seen in the field. In the following discussion, we take our data at face value and assume that the C/M ratios measured in Fields 1 and 5 are indeed the true C/M ratios.

In Sec. 3.4 we derived AGB LFs for the five fields and showed that significant differences existed between them. Comparing these (Fig. 9) with the C/M ratios in Fig. 13 we see no simple correlation. For example, Field 3 has the shallowest AGB LF and a strong blue component, indicative of ongoing star formation, though has a lower C/M ratio than Field 4. We conclude that, whereas we cannot rule star-forming history out as playing a role in determining the C/M ratio, it is not the dominant factor in M31. Consequently, we can say that age, and hence mass, is not the predominant factor that drives the C/M ratio in M31.

Simple inspection of Fig. 10 reveals that the C/M ratio increases with $R_{\rm M31}$. This is as expected if C/M correlates with metallicity and if metallicity in M31 is inversely correlated with $R_{\rm M31}$. An anticorrelation between metallicity and galactocentric distance is seen in many spiral galaxies (see Dinerstein 1990 for a review) and has been observed in M31.

Before pursuing the above idea further, we first dismiss an alternative interpretation of the data in Fig. 13. The data in Fig. 13 could be construed as showing that there are two values for the C/M ratio in M31, one for the bulge (inner two data) and one for the disk (outer three data). We note however that the effective radius of the bulge of M31 is around 2 kpc (Walterbos & Kennicutt 1988) whereas our inner two fields lay at $R_{\rm M31} = 4.6$ and 7.2 kpc respectively.

The metallicity gradient in M31 has been investigated by Dennefeld & Kunth (1981) and Blair, Kirshner & Chevalier (1981, 1982). As pointed out by Blair, Kirshner & Chevalier (1982, henceforth BKC82) M31 has often been neglected for abundance studies due to its inclination and relative lack of large, bright, H II regions. In the most recent study (BKC82) abundances were derived using both supernova remnants (SNRs) and H II regions. BKC82 measured a consistent gradient for N using both SNRs and H II regions, but found that whereas H II regions showed an O gradient, SNRs did not. BKC82 attributed the inconsistency to problems with the SNR shock model. This result aside, they found that in M31 both N and O abundances decreased by a factor of 4 from approximately 4 to 23 kpc.

How well do estimates of the metallicity in our fields (derived from the C/M ratios) compare with the (O/H) measurements of BKC82? We obtain estimates for [m/H] in our fields from an unweighted least-squares fit to Fig. 7 of Pritchet *et al.* (1987), which we show in Fig. 17. This gives us the expression

$$[{\rm m/H}] = -0.53 \log({\rm C/M}) - 0.49 \tag{3}$$

for [m/H] in terms of the C/M ratio. Assuming that in our fields [m/H] = [O/H] (note our nomenclature: (O/H) is the number ratio, while [O/H] is the logarithm of the abundance relative to solar values), and adopting a value of $6.6\times10^{-4}$ for the solar number ratio of (O/H) (Allen 1976), we have

$$({\rm O/H}) = 6.6 \times 10^{-4} (10^{[{\rm m/H}]}) \tag{4}$$

for (O/H) in our fields. In Fig. 18 we show a comparison between the (O/H) gradient derived by BKC82 and that using the C/M ratio as derived above. As can be seen, reasonable agreement is seen up to the limit of the BKC82 data. Beyond the limit of the BKC82 data, *the Field 5 point suggests that the metallicity gradient flattens out.* Indeed, the BKC82 data does show evidence of flattening for $R_{\rm M31} > 15$kpc. We strengthen the suggestion made by RCP90 that the metallicity gradient in M31 flattens out.

The above result is of importance for two reasons. Firstly, the metallicity determined from the C/M ratio tells us about the metallicity gradient in the disk of M31 a few Gyr ago when the AGB stars formed, in contrast to H II regions which tell us about abundances at the present epoch. The agreement between (O/H) ratios derived from H II regions and those from C/M ratios implies that no significant chemical evolution has occurred in the disk of M31 during this time. As discussed by Peimbert (1995), planetary nebulae can also provide us with information about abundance gradients in the past. Confirmation that the abundance gradient in M31 has remained relatively unchanged over the last few Gyr, could be obtained from a study of the planetary nebulae in M31.

Secondly, knowledge of the abundance gradient can be used to constrain models of galaxy formation. For example, using a star-forming viscous disk model for spiral galaxy disks, Zaritsky (1992) predicts that the radial abundance profile of a noninteracting unbarred spiral galaxy changes slope at a radius where the rotation curve changes from linearly rising to flat. In M31 this happens at $30'$-$40'$ (Rubin & Ford 1970) which corresponds to $R_{\rm M31} = 6$ to 8 kpc. The steep rise in the C/M ratio from Field 2 to Field 3 and the slower rise from Field 3 to Field 4 provides tantalizing evidence that we may be seeing a manifestation of such an effect in M31.

Turning the above arguments around we can use the BKC82 data to obtain metallicity estimates for our fields and see how well our data fit the trend from Pritchet *et al.* (1987) shown in Fig. 17. In Fig. 6 of BKC82 a fit is given to the (O/H) ratio measured in M31 which we include in our reproduction of this figure in Fig. 18. A comparison with the cosmic (O/H) ratio (Allen 1976) shows that at $R_{\rm M31} = 15$ kpc the abundance is approximately solar. Using Fig. 6 of BKC82 we derive an abundance gradient of –0.035 dex/kpc. We can thus give the abundance in the disk of M31, relative to solar, as

$$[{\rm O/H}] = 0.035(15 - R_{\rm M31}) \tag{5}$$

for $23 > R_{\rm M31} > 4$ kpc. In Fig. 19 we plot data from Table 5 of Pritchet *et al.* (1987) as well as points for our fields. In Fig. 19 we include the point for Field 5 (despite the fact that the metallicity determination was derived by extrapolation) to demonstrate that the Field 5 point is an outlier if it is assumed the metallicity gradient in the disk of M31 does not flatten. Indeed, if the abundance gradient in M31 continues out to Field 5, then we might expect Field 5 to have a C/M ratio about ten times greater than that seen. As pointed out by Pritchet *et al.* (1987), the C/M ratios in their plot may not be strictly comparable due to the varying definitions of M-stars and metallicity. Nonetheless, Fig. 19 clearly shows that the new, internally consistent, measurements of C/M ratio in M31 follow the relationship that is seen to hold over 4 dex in (C/M) and 1.5 dex in [m/H] for the many types of galaxies in Fig. 19. *From this we conclude that it is metallicity and not galactic morphological type that determines the C/M ratio.*

### *4.3.5. Reasons for the correlation.*

Why might we expect the C/M ratio to be correlated with metallicity? The C/M ratio data from this and other surveys can be explained if either: (1) the number of M-stars decreases with metallicity; (2) the number of C-stars increases with metallicity; (3) a combination of (1) or (2). A physical reason for the first of these possibilities is that in a metal-poor environment the color of the AGB will be shifted bluewards and hence fewer stars will appear as M-types. Under this hypothesis it would be expected that the number of C-stars scales as the luminosity of the field, a hypothesis ruled out by the data from Fields 1 and 2. This suggests that whereas we cannot rule out the color of the AGB as playing a role in determining the C/M ratio, we must look to the second possibility, an increase in the number of C-stars with decreasing metallicity, for an explanation of the correlation.

From our data it appears that the C-stars are what we term *mepephiles* (lovers of MEtal-Poor Environments). This may be because it is easier to turn a metal-poor AGB star into a C-star, since (1) a metal-poor star will have a less extended envelope and so convection does not need to reach so deep for the star to undergo third dredge-up, and (2), in a metal-poor star, less dredge-up is needed to drive C>O. A competing explanation for why C-stars are mepephiles is that the metal-rich C-stars have higher mass-loss rates, and are short lived compared to their metal-poor counterparts, or evolve into infrared C-stars which are undetected in this survey. Theory should help guide us here. For example, if it can be shown that the time for third dredge-up to turn a metal-rich AGB star into a C-star is only slightly longer than the time to turn a metal-poor AGB star into a C-star compared with the lifetime of these C-stars, then the mass-loss explanation is more likely. Two observational studies which may help cast light on the matter are: (1) an investigation of how the ratio of carbon- to oxygen-rich planetary nebulae (PNe) varies with $R_{\rm M31}$. If the C/O PNe ratio tracks the C/M star ratio then the variation is a metallicity effect, whereas if the ratios do not track each other (for example the C/O PNe ratio remains constant) then other causes such as lifetimes are involved in determining the C/M ratio. An extensive catalog of PNe in M31 already exists (Ford & Jacoby 1978), although obtaining carbon abundances for PNe is difficult, requiring ultraviolet spectra and HST observations (Jacoby 1994); (2) A search for IR C-stars in our fields. Frogel & Richer (1983) found there to be a lack of luminous IR C-stars in a LMC field, suggesting that luminous AGB stars undergo high mass-loss rates and are consequently short lived. A similar study in M31 should allow the role of metallicity in mass-loss rates to be better understood. For example, does the C/M ratio remain unchanged with $R_{\rm M31}$ if IR C-stars are included?

## 5. THE C/M RATIO AND THE ISM

Using IRAS data, Thronson *et al.* (1987) showed that the highly-evolved carbon-rich and oxygen-rich stars in the Galaxy have different surface densities as a function of Galactocentric distance (see their Fig. 6). The sense of this relationship is an increasing ratio of carbon-rich to oxygen-rich stars with Galactocentric distance, in the same sense as the variation of the C/M ratio in M31. Thronson *et al.* (1987) note that oxygen-rich stars will be silicate-grain generators whereas carbon-rich stars will produce carbonaceous grains. They note that though the change in elemental abundances may be modest (depending on the poorly understood roles of novae and supernovae), the grain composition will have a strong radial variation in the Galaxy. Thronson *et al.* (1987) mention three areas in which the varying grain composition might be expected to have an effect. Briefly, these are

- temperature regulation of dense molecular clouds. This will affect star formation and lead to the IMF being position-dependent.
- extinction laws. As the grain composition varies as a function of position in the Galaxy, there would be no universal interstellar extinction law.
- molecular cloud chemistry.

To these we add another area in which we believe grain composition may have an effect:

- The ratio of diffuse interstellar band (DIB) strength to interstellar reddening. Such a correlation may arise if DIBs are caused by carbonaceous grains, which we expect to be more dominant at greater galactocentric distances in both M31 and the Milky Way.

We have shown that the C/M ratio varies in M31, and so we expect that the phenomena described above may be present in M31. In Sec. 5.1 we cite evidence in favor of a connection between the C/M ratio and extinction laws in M31.

## 5.1. Effects on Extinction

As noted by IR83, the different proportions of C-stars in the SMC, LMC and Milky Way should translate into different relative proportions of carbonaceous- and silicate-grains in the interstellar medium (ISM) of these galaxies. They mention that the differences between the UV extinction curves of these three galaxies correlate with their C-star abundances. If extinction laws are affected by the grain composition, as this suggests, then a correlation between C/M ratio and extinction laws would be expected. In M31, the radial dependence of the C/M ratio would lead to the extinction law varying with galactocentric distance. (Caveat emptor: Cardelli (1994) attributes deviations in ultraviolet extinction laws to changes in the ISM such as grain growth, destruction and changes in grain size distribution and not to variations in composition).

Evidence that the extinction law in M31 varies was reported by L. Searle in the Carnegie Year Book (1982, pp. 622-624). Searle studied globular clusters in M31 and, using reddening free parameters, showed that the reddening law varied systematically with galactocentric distance. Searle continued his study of dust in M31 along with I. Thompson and other collaborators. A brief report on the progress of their investigations is given in the Carnegie Year Book (1985, pp. 68-69). In this report they state that the extinction in the violet (for a given $(V-K)$ reddening) decreases radially outward. In this report the following speculation is given as to the origin of this relationship:

> "The dust responsible for interstellar reddening is thought to have been formed in, and then ejected from, the atmospheres of cool evolved giant stars. There are two common types of such stars, Carbon stars and M stars, whose atmospheres contain free carbon and free oxygen respectively. In our Galaxy, the ratio of these two types of stars is known to vary systematically with distance from the center of the Galaxy. It seems likely that the radial variation in the dust properties of M31 has its origin in a similar change in the populations of dust producing stars."

Before this study it was known that both the extinction law in M31, and the C/M ratio in our Galaxy, varied with galactocentric distance. We can start to pull the evidence in favor of a connection between the C/M ratio and extinction laws together by now stating that in M31 *both the extinction law and C/M ratio vary with galactocentric distance.* As mentioned by L. Searle (Carnegie Yearbook, 1982), it is no surprise that an extinction law gradient in the Galaxy has escaped detection as remote regions in the plane of the Galaxy are heavily obscured and cannot easily be observed. Observational difficulties aside, it would be of great interest to demonstrate that the Galactic extinction law behaves in the same way as that of M31.

In their investigation of M31 Cepheids, the variation in M31's extinction law was discussed by FM90. This was of concern to them as an incorrect extinction law leads to an incorrect distance modulus for their Cepheids. Such an error would ultimately be propagated into measurements of the Hubble constant. FM90 mention that the sense of Searle's measured color excesses is such that the ratio of $A_V/E_{B-V}$ would be expected to increase radially outwards, and that such a trend is consistent with the findings of van den Bergh (1968) and Kron & Mayall (1960). Using realistic estimates of $A_V/E_{B-V}$, FM90 found the effects of a varying extinction law to lie within the noise of their data.

How might we expect the C/M ratio to affect the extinction laws? Thronson *et al.* (1987) mention that a common description of interstellar extinction has carbonaceous grains dominating visual and near-infrared extinction, and silicate grains, with their large mantles, dominating mid- and far-infrared extinction (Draine & Lee 1983). The picture then makes sense in as much as FM90 pointed out that the ratio of $A_V/E_{B-V}$ increases radially outward, as would be expected there were an increasing number of carbonaceous grains at greater radial distances. The increased number of carbonaceous grains is a natural consequence of the increase in the ratio of C- to M-stars at greater radial distances.

It would be of great interest to investigate whether M31's extinction law remains unchanged beyond where the C/M ratio has flattened out. If the extinction law was found to track the C/M ratio, this would be strong evidence in favor of the C/M ratio determining the grain composition, and the grain composition determining the extinction law.

## 6. CONCLUSIONS

We have used four-band photometry to investigate 5 fields along the SW major axis of M31. The main findings are as follows.

1. A distance modulus to M31 ($(m-M)_0 = 24.38$) is derived, using previously identified Cepheids in Field 3. The value derived is consistent with other values in the literature.
2. The $(V-I)$ color width of the GB is shows differences between the fields.
    - The GB width is inconsistent with (realistic) estimates of the metallicity dispersion.

- Much of the GB width is consistent with stars of varying masses funnelling into the GB region of the CMD.
- The contribution from differential reddening remains unknown.

3. The five fields have different AGB LFs, showing that they have distinct star-forming histories.

4. A distance modulus derived using C-stars in Fields 2, 3 and 4 ($(m-M)_0 = 24.36$) is in good agreement with the distance modulus derived from the Cepheids. The mean C-star magnitude was consistent between these three fields, despite differences in their star-forming history and metallicity.

5. The ratio of C-stars to M-stars was shown to increase with galactocentric distance in M31. Completeness tests on the data showed this result to be independent of different crowding conditions in the fields.

6. Under the assumption that the C/M ratio is a metallicity indicator, we derived metallicity values for our fields using previous observations as a guide. We then showed that there was good agreement between our data and measurements from H II regions and that our data indicate a flattening of the metallicity gradient in M31 beyond $\sim$ 15 kpc.

7. We considered the impact that a varying C/M ratio would have on the ISM. The main conclusions that we arrived at here were that there is evidence that the extinction law in M31 is compatible with what would be caused by the observed C/M ratio gradient, and that a fruitful area of future research may be to look for a correlation between the ratio of DIB strength and reddening against Galactocentric distance.

JPB wishes to thank Brad Gibson and Jaymie Matthews for useful discussions during the preparation of this paper. The research discussed in this paper was supported by grants to HBR from the Natural Sciences and Engineering Research Council of Canada. JPB is indebted to UBC for providing a University Graduate Fellowship for the years 1991-94.

## REFERENCES

Aaronson, M., Blanco, V.M., Cook, K.H., & Schechter, P.L. 1989, ApJS, 70, 637

Aaronson, M., Da Costa, G. S., Hartigan, P., Mould, J.R., Norris, J., & Stockman, H.S. 1984, ApJ, 277, L9

Aaronson, M., & Mould, J. 1982, ApJS, 48, 161

Allen, C.W. 1976, *Astrophysical Quantities* (London, Athlone Press)

Baade, W. & Swope, H. H. 1965, AJ, 70, 212

Bertelli, G., Bressan, A., Chiosi, C., Fagotto, F. & Nasi, E. 1994, A&AS, 106, 275

Bessell, M.S. 1979, PASP, 91, 589

Bessell, M.S., & Brett, J.M. 1988, PASP, 100, 1134

Bessell, M.S. & Wood, P.R. 1984, PASP, 96, 247

Blair, W.P., Kirshner, R.P. 1981, ApJ, 247, 879

Blair, W.P., Kirshner, R.P. & Chevalier, R.A. 1982, ApJ, 254, 50 (BKC82)

Blanco, B.M., Blanco, V.M., & McCarthy, M.F. 1978, Nature, 271, 638

Blanco, V.M, McCarthy, M.F., & Blanco, B.M. 1980, ApJ, 242, 938

Bolte, M., 1989, ApJ, 341, 168

Brewer, J.P., Fahlman, G.G., Richer, H.B., Searle, L., & Thompson, I. 1993, AJ, 105, 2158

Brewer, J.P., Crabtree, D.R, & Richer, H.B. 1995, *In preparation.*

Burstein, P., & Heiles, C. 1984, ApJS, 54, 33

Cameron, L.M. 1987, A&A, 147, 39

Cardelli, J. 1994, Private Communication

Cohen, J.G., Frogel, J.A., Persson. S.E. & Elias, J.H. 1981, ApJ, 249, 481

Cook, K.H., Aaronson, M., & Norris, J. 1986, ApJ, 305, 634

Da Costa, G.S, & Armandroff, T.E. 1990, AJ, 100, 162

Davis, L. 1994, Private Communication

Dennefeld, M., & Kunth, D. 1981, AJ, 86, 989

de Vaucouleurs, G. 1958, ApJ, 128, 465

Dinerstein, Harriet L. 1990, in *The Interstellar Medium in Galaxies* edited by Thronson & Shull (Holland:Kluwer)

Draine, B.T., & Lee, H.M. 1984, ApJ, 285, 89

Drukier, G.A., Fahlman, G.G., Richer, H.B., & VandenBerg, D.A. 1988, AJ, 95, 1415

Ford, H.C. & Jacoby, G.H. 1978, ApJS, 38, 351

Freedman, W.L., Hughes, S.M., Madore, B.F., Mould, J.R., Lee, M.G., Stetson, P., Kennicutt, R.C., Turner, A., Ferrarese, L., Ford, H., Graham, J.A., Hill, R., Hoessel, J.G., Huchra, J., & Illingworth, G.D. 1994, ApJ, 427, 628

Freedman, W.L., & Madore, B.F. 1990, ApJ, 365, 186 (FM90)

Frogel, J.A., & Richer, H.B. 1983, ApJ, 275, 84

Gaposchkin, S. 1962, AJ, 67, 334

Green, P.J. 1992, PASP, 104, 977

Green, P.J., Margon, B., Anderson, S.F., & Cook, K.H. 1994, ApJ, 434, 319

Hodge, P. 1979, AJ, 84, 744

Hodge, P. 1981, *Atlas of the Andromeda Galaxy* (Seattle: University of Washington Press)

Hodge, P., & Lee, M.G. 1988, ApJ, 329, 651

Hudon, J.D., Richer, H.B., Pritchet, C.J., Crabtree, D, Christian, C.A. & Jones, J. 1989, AJ, 98, 1265

Iben, I. Jr., 1975, ApJ, 196, 525

Iben, I. Jr., 1981, ApJ, 246, 278

Iben, I. Jr., & Renzini, A. 1983, ARA&A, 21, 271 (IR83)

Iben, I. Jr., & Truran, J.W. 1978, ApJ, 220, 980

Iye, M., & Richter, O.G. 1985, A&A, 144, 471

Jacoby, G.H. 1994, Private Communication

Knapp, G.R. 1991, in Frontiers of Stellar Evolution, edited by D.L. Lambert (ASP Conf. Ser., 20), 229

Kraft, R.P. 1963, in *Stars and Stellar Systems,* Vol. 3. *Basic Astronomical Data,* edited by K.A. Strand (Chicago: University of Chicago Press), p. 421

Krisciunas, K. 1994, Private Communication

Kron, G.E., & Mayall, N.U. 1960, AJ, 65, 581

Landolt, A.U. 1992, AJ, 104, 340

Lee, M.G., Freedman, W.L., & Madore, B.F. 1993, ApJ, 417, 553

Lilly, S.J. 1987, MNRAS, 229, 589

Madore, B.F., & Freedman, W.L. 1991, PASP, 103, 933 (MF91)

Mihalas, D. & Binney, J. 1981, *Galactic Astronomy* (San Francisco: W.H. Freeman and Company)

Palmer, L.G., & Wing, R.F. 1982, AJ, 87, 1739

Peimbert, M., 1995, in Highlights in Astronomy, Vol. 10, edited by I. Appenzeller (Dordrecht:Kluwer)

Pierce, M.J., Welch, D.L., McClure, R.D., van den Bergh, S., Racine, R., & Stetson, P.B. 1994, Nature, 371, 385

Pritchet, C.J., Richer, H.B., Schade, D., Crabtree, D.R. & Yee, H.K.C 1987, ApJ, 323, 79

Reid, N. & Mould, J. 1984, ApJ, 284, 98

Renzini, A. 1977, in Advanced Stages of Stellar Evolutions, edited by P. Bouvier & A. Maeder (Geneva: Geneva Observatory), p. 149

Renzini, A. 1981, Ann. Phys. (Paris) 6, 87

Renzini, A., & Voli, M. 1981, A&A, 94, 175

Richer, H.B. 1981, ApJ, 243, 744

Richer, H.B. & Crabtree, D.R. 1985, ApJ, 298, L13 (RC85)

Richer, H.B., Crabtree, D.R. & Pritchet, C.J. 1984, ApJ, 287, 138 (RCP84)

Richer, H.B., Crabtree, D.R. & Pritchet, C.J. 1990, ApJ, 355, 448 (RCP90)

Richer, H.B., Pritchet, C.J. & Crabtree, D.R. 1985, ApJ, 298, 240

Rubin, V. C. & Ford, W.K. 1970, ApJ, 159, 379

Schwarzschild, M. & Härm, R. 1965, ApJ, 142, 855

Stetson, P.B. 1990, PASP, 102, 932

Stetson, P.B. 1992, in Astronomical Data Analysis Software and Systems I, ASP Conference Series, Vol. 25, edited by D. M. Worrall, C. Biemesderfer, and J. Barnes (ASP, San Francisco), p. 297

Stetson, P.B. 1993, in Stellar Photometry - Current Techniques and Future Developments, IAU Coll. 136, edited by C. J. Butler and I. Elliot (Cambridge: Cambridge University Press)

Stetson, P.B., Davis, L.E., & Crabtree, D.R. 1990, in CCDs in Astronomy, ASP Conference Series, Vol. 8, edited by G.H. Jacoby (ASP, San Francisco), p. 289

Stetson, P.B., & Harris, W.E. 1988, AJ, 96, 909

Taylor, B.J. 1986, ApJS, 60, 577

Thronson, H.A., Latter, W.B., Black, J.H., Bally, J., & Hacking, P. 1987, ApJ, 322, 770

van den Bergh, S. 1968, *Observatory*, 88, 168

Walterbros, R.A.M., & Kennicutt, R.C. Jr. 1988, A&A, 198, 61

Welch, D.L., McAlary, C.W., McLaren, R.A., & Madore, B.F. 1986, ApJ, 305, 583

Zaritsky, D. 1992, ApJ, 390, L73

Fig. 1.— A montage of 2 Palomar Sky Survey E-plates showing the positions of our five fields in M31. The fields are approximately contained within the square outlines. The scale bar indicates a length of 10 kpc at a distance of 770 kpc, the assumed distance to M31. SUPPLIED ON REQUEST.

Fig. 2.— The $I$, $(V-I)$ CMDs for the five fields in M31. The data are calibrated, although no reddening corrections have been applied. In the lower right panel parallel dashed lines show the region occupied by Galactic Cepheids, while the redder solid line is the 47 Tuc fiducial and the bluer solid line the M15 fiducial The upper sequence of points shows the positions of supergiants with spectral types M0Ib, M2Ib, M3Ib and M4Ib, while the lower sequence of points shows the positions of giants with spectral types M0II, M2II, M3II and M4II. These features have been located in the CMD as described in Sec. 3. SUPPLIED ON REQUEST.

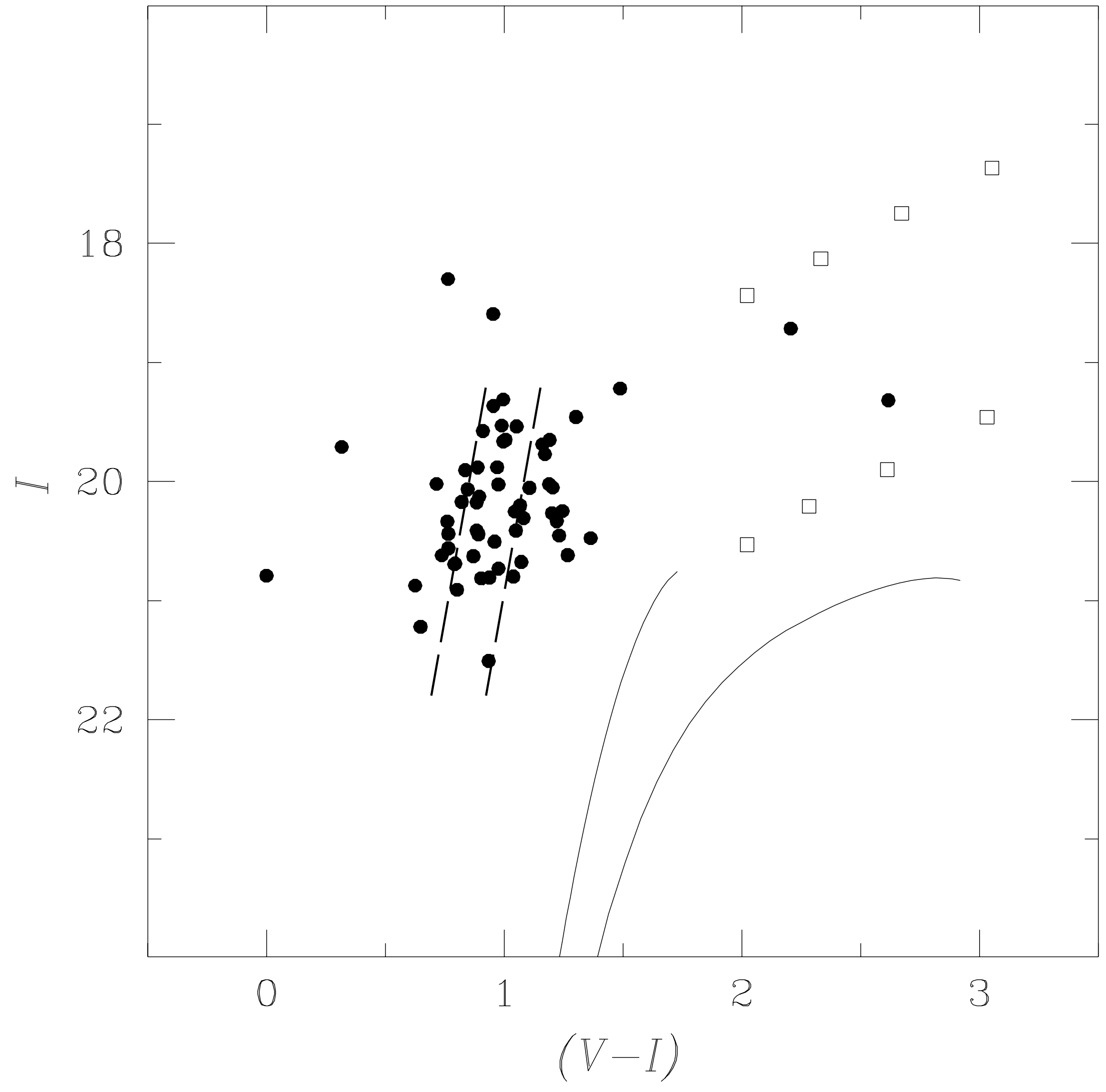

Fig. 3.— An $I$, $(V-I)$ CMD for the Cepheids from Baade & Swope (1965) that were reidentified in Field 3. We include in the CMD the features shown in the bottom right panel of Fig. 2 (see Sec. 3 and the Fig. 2 caption). Of special note is the expected position of the Cepheid instability strip, which is shown by parallel dashed lines.

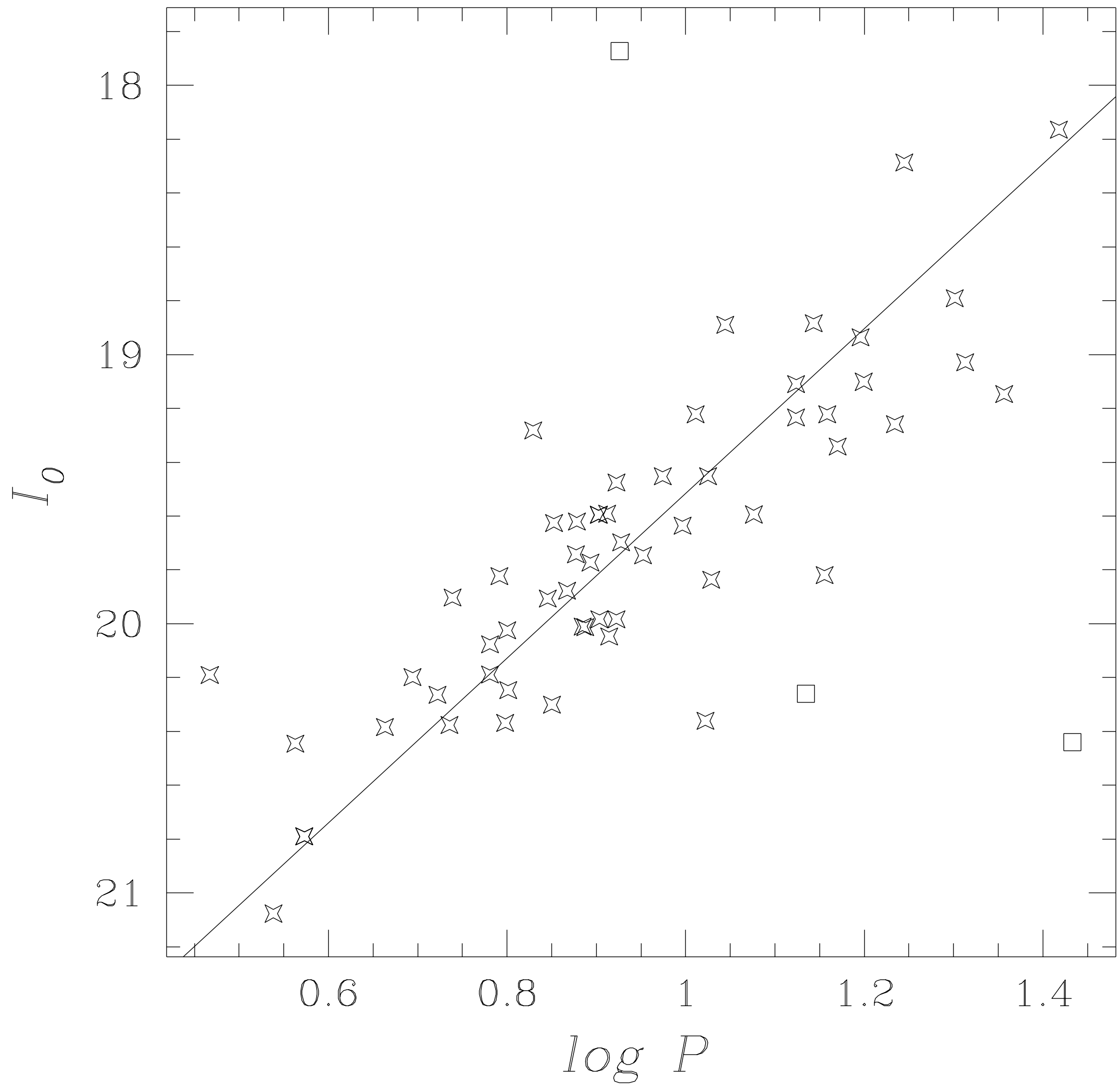

Fig. 4.— Unphased P-L diagram for Cepheids in Field 3. The periods are from Baade & Swope (1965), while the $I$ magnitudes, corrected for absorption, are from this study. The solid line is a fit of the P-L relationship of MF91 given in Eq. (1). Open squares indicate those stars whose $I$ magnitude was more than 2 standard deviations away from the initial fit. The errors in the $I$ magnitudes returned by ALLSTAR were around 0.02 for the brighter Cepheids and 0.04 for the fainter Cepheids.

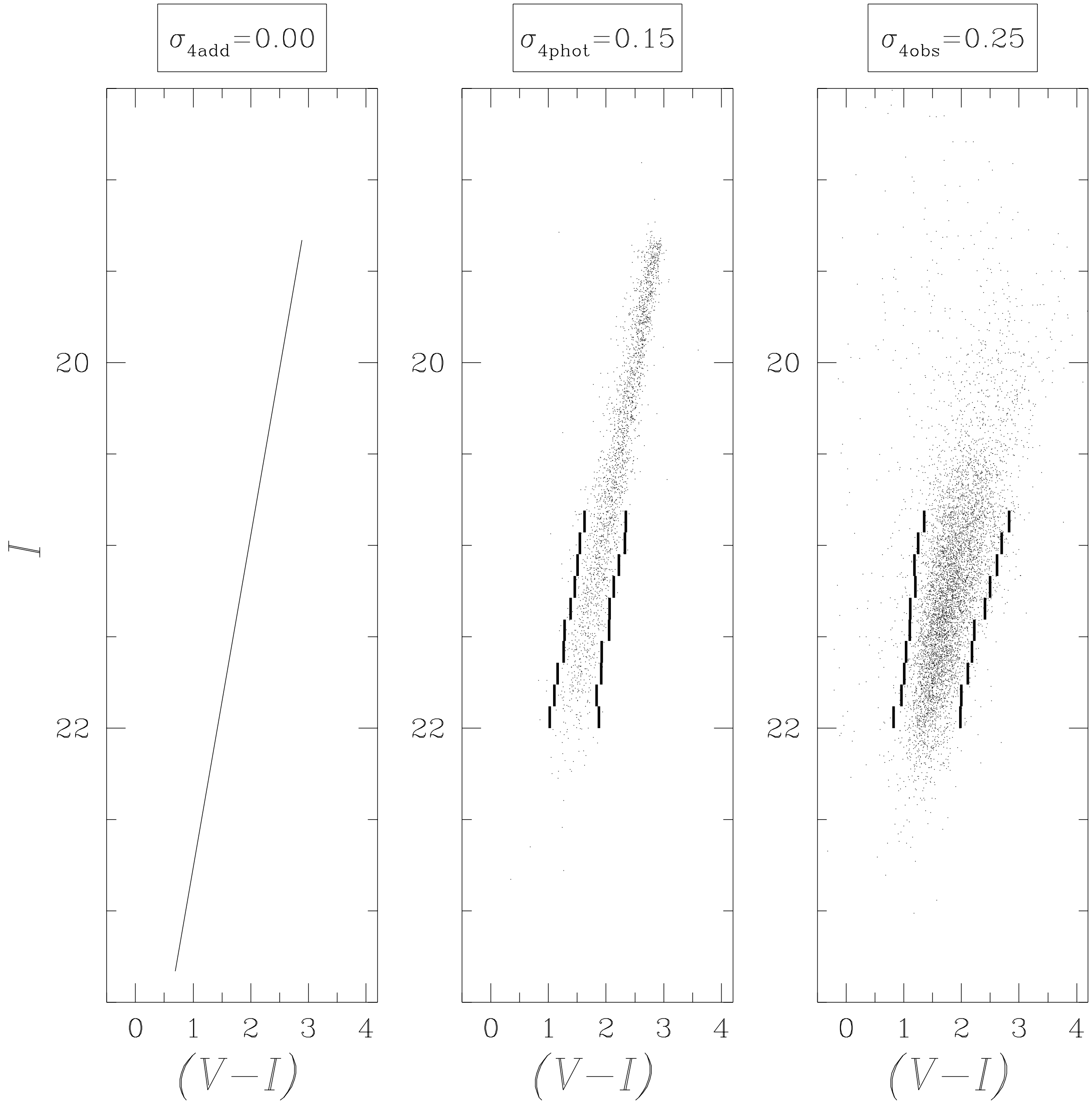

Fig. 5.— The leftmost panel shows the dispersionless relation used to generate artificial stars which were added to the Field 4 $V$ and $I$ frames. The middle panel shows the artificial stars which were recovered and matched from the $V$ and $I$ frames, while the rightmost panel shows real data. The stars contained between the vertical lines in the right two panels are the stars from which $\sigma$ was measured.

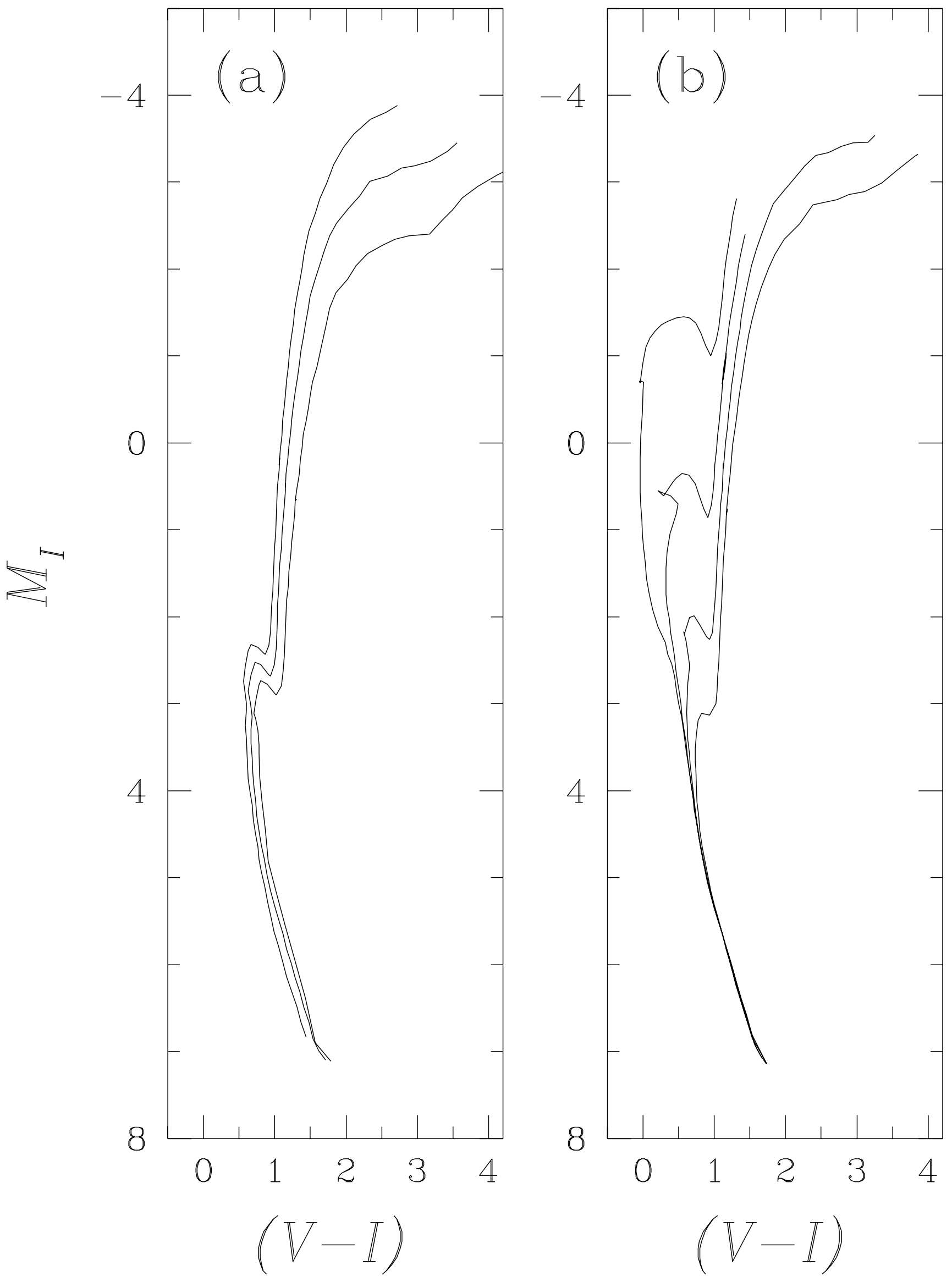

Fig. 6.— Isochrones from Bertelli *et al.* (1994). In the left panel we plot 5 Gyr isochrones for Z values (from left to right) of 0.008, 0.02, and 0.05. The other panel shows isochrones all with Z=0.02, but with ages (from left to right) of $3.2\times10^8$, $1\times10^9$, $3.2\times10^9$, and $1\times10^{10}$ years.

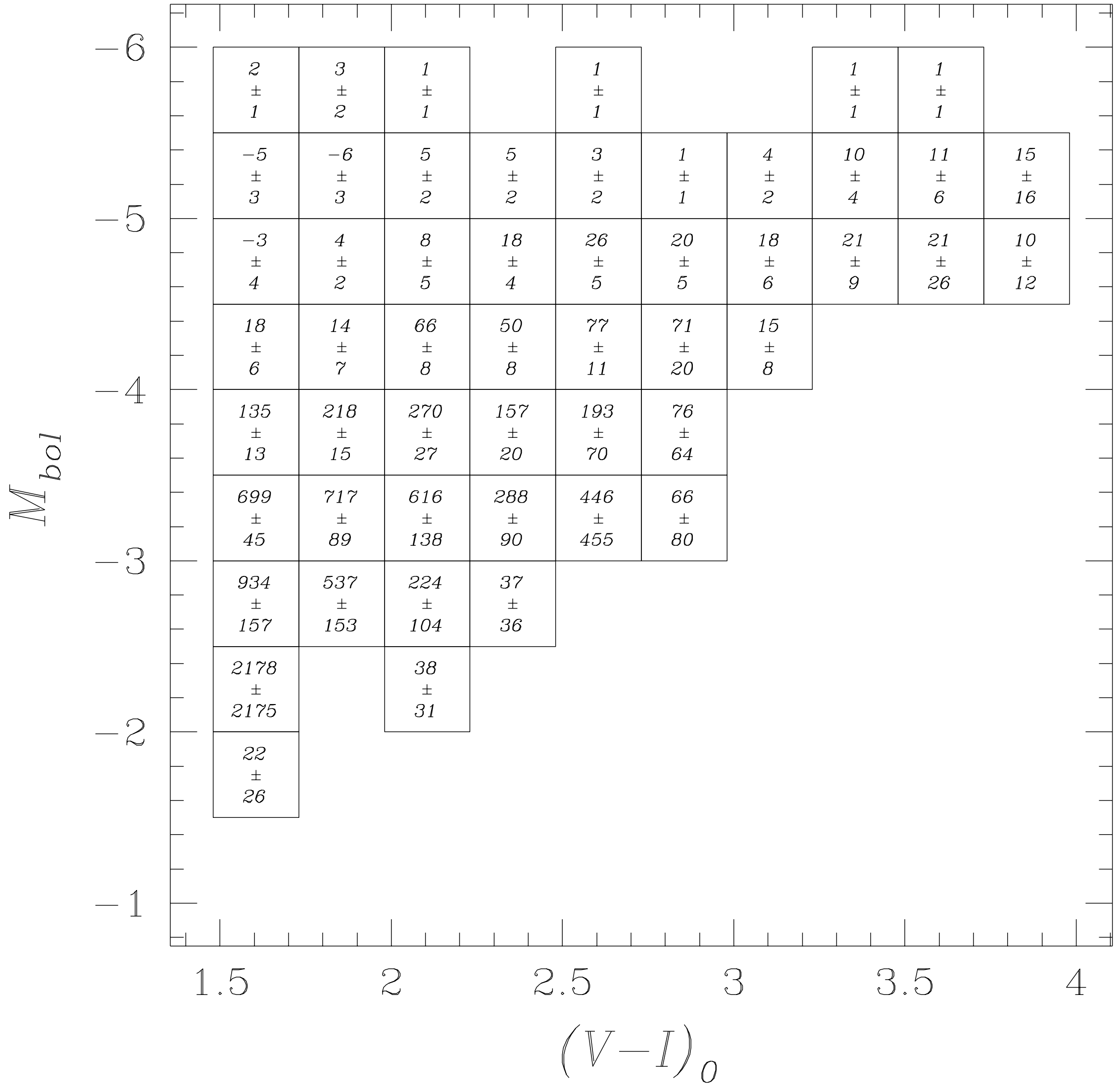

Fig. 7.— This Figure shows the corrected counts for Field 4 stars on the $M_{\rm bol}$, $(V-I)_0$ plane. Stars were added to the plane in a 10×10 grid, though bins are only shown if real stars existed in the bins and completeness corrections were finite (at least one added star was recovered). The upper number in the bins is the completeness and background corrected count, while the lower number is the error in the corrected count.

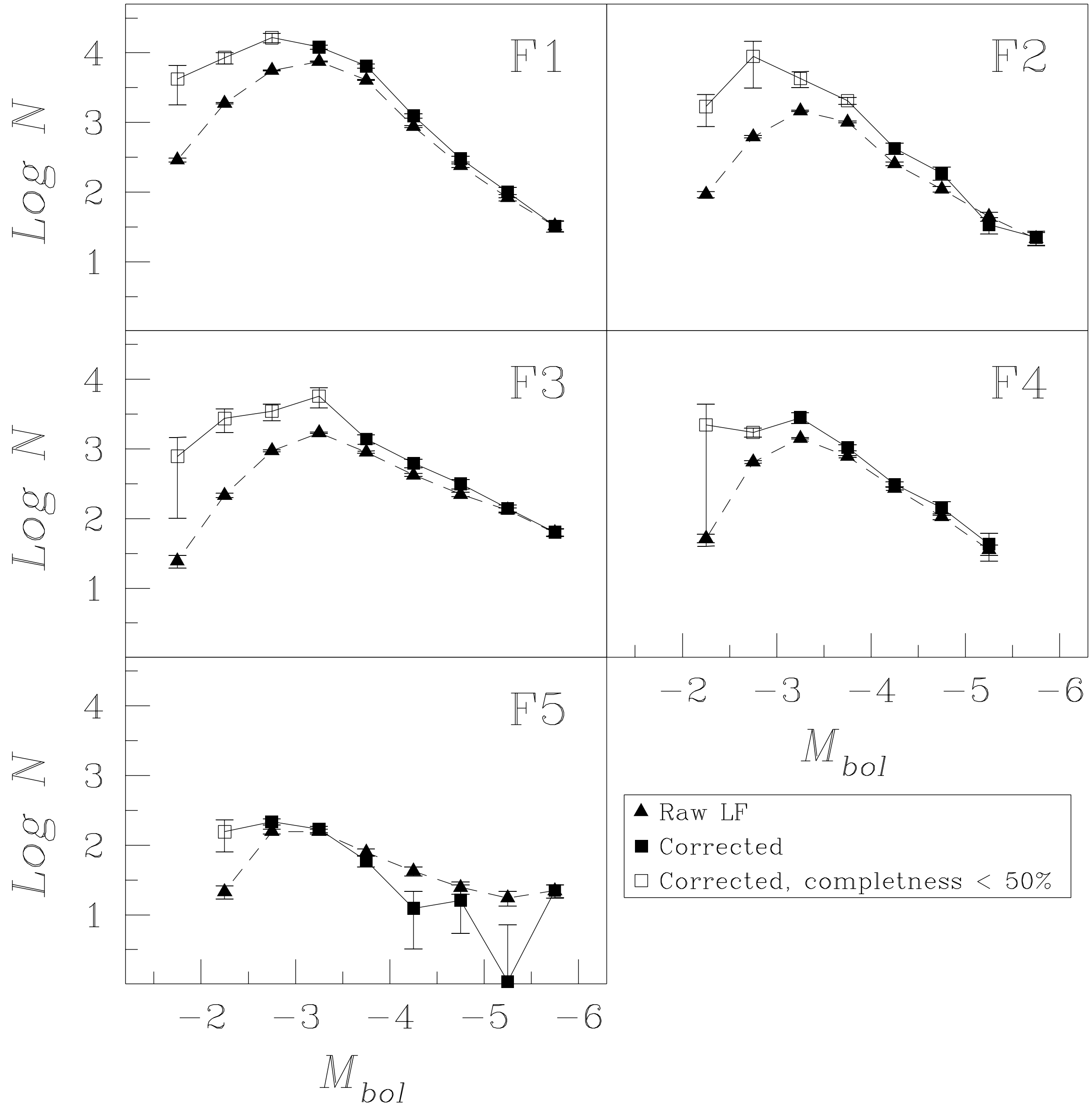

Fig. 8.— In the above diagram bolometric LFs for stars with $1.48 < (V-I)_0 < 3.98$ in the five M31 fields are shown. Solid triangles indicate the LFs before corrections for background stars and completeness, whereas squares indicate the LFs after correction for these two effects. An open square indicates that the applied completeness correction was greater than 2.

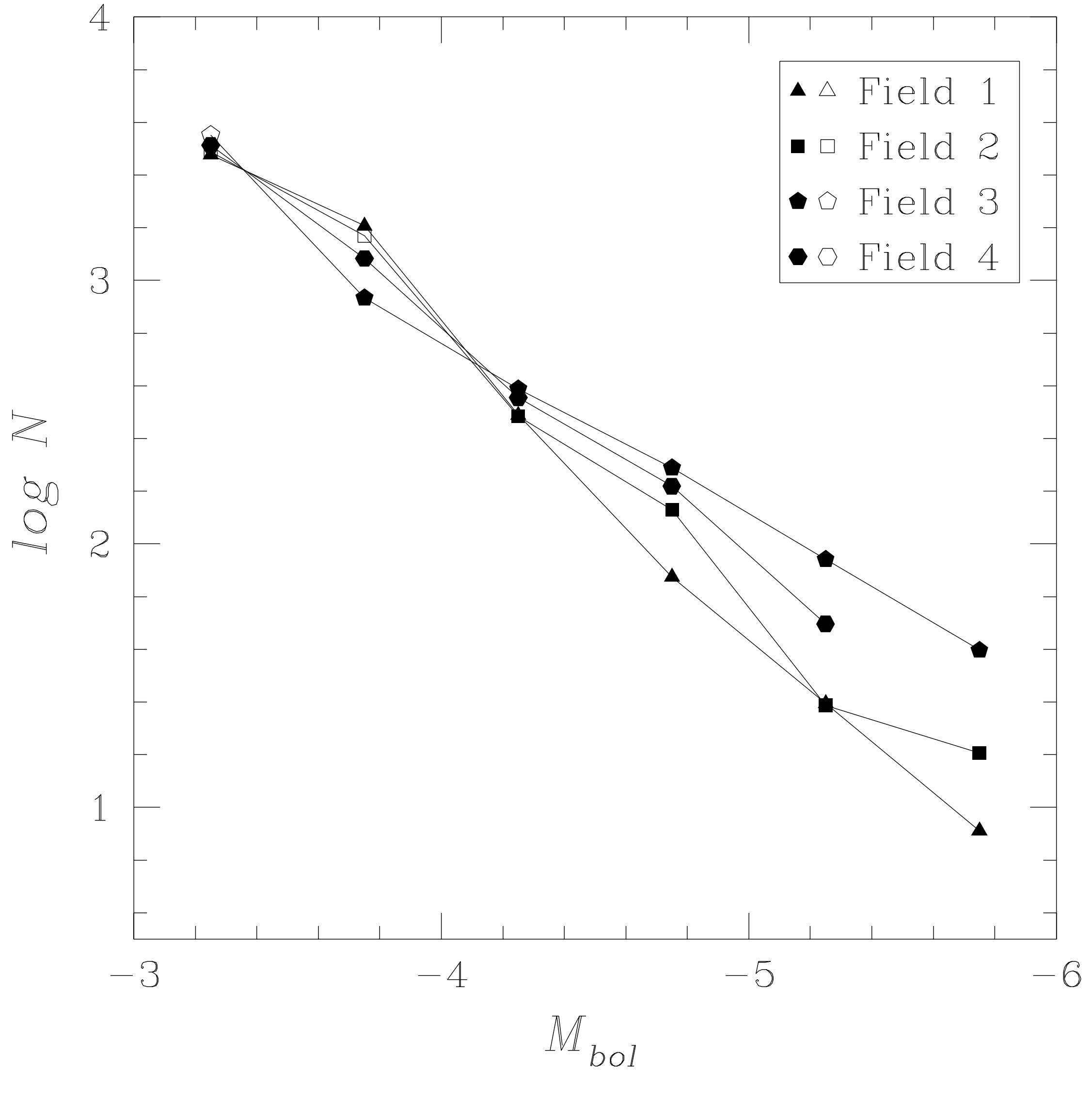

Fig. 9.— This plot contains the AGB LFs for Fields 1 through 4 from Fig. 8. The LFs were normalised such that they all contained 5000 stars in the magnitude range $-5 < M_{\rm bol} < -3$.

Fig. 10.— The labeled panels show the $(CN-TiO,\ V-I)$ color-color diagrams for our five fields in M31. The $(V-I)$ color has been calibrated onto the Johnson-Cousins system and corrected for reddening (Sec. 3.1), while the $(CN-TiO)$ color has been shifted so the mean value of $(CN-TiO)$ for stars bluer than $(V-I) = 1$ is zero. The number in the lower left corner of the panels indicates the number of stars plotted in each diagram. The lower right panel shows the boundaries used to define the C- and M-stars, as discussed in the text (Secs. 4.1 and 4.3.2). SUPPLIED ON REQUEST.

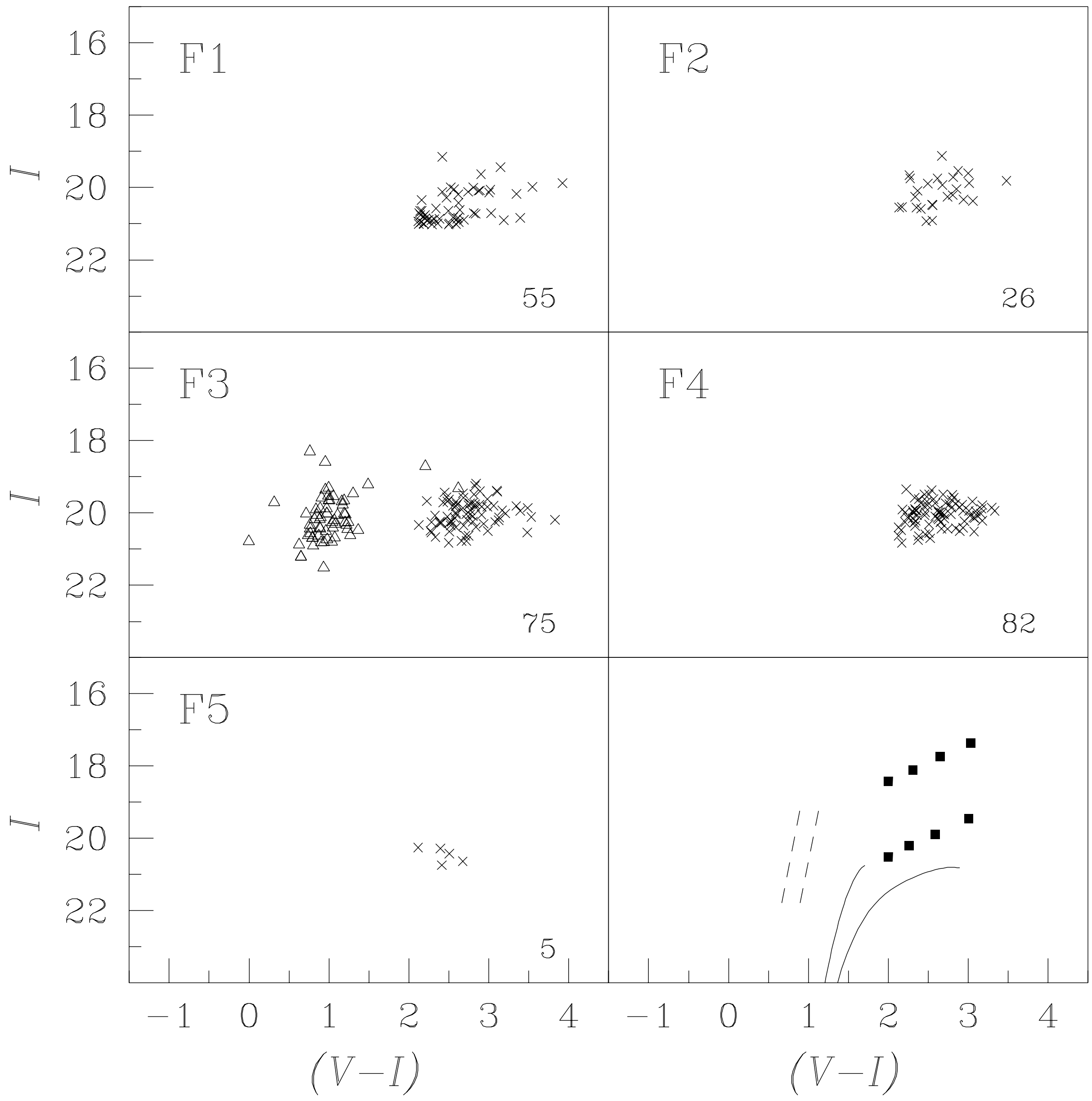

Fig. 11.— Calibrated $I$, $(V-I)$ CMDs for the C-stars identified in M31. The C-stars (shown as crosses) were selected according to the color and magnitude criteria described in Sec. 4.1. The bottom right panel shows features described in the Fig. 2 caption and Sec. 3. The Baade & Swope (1965) Cepheids which we reidentified in Field 3 are shown as triangles.

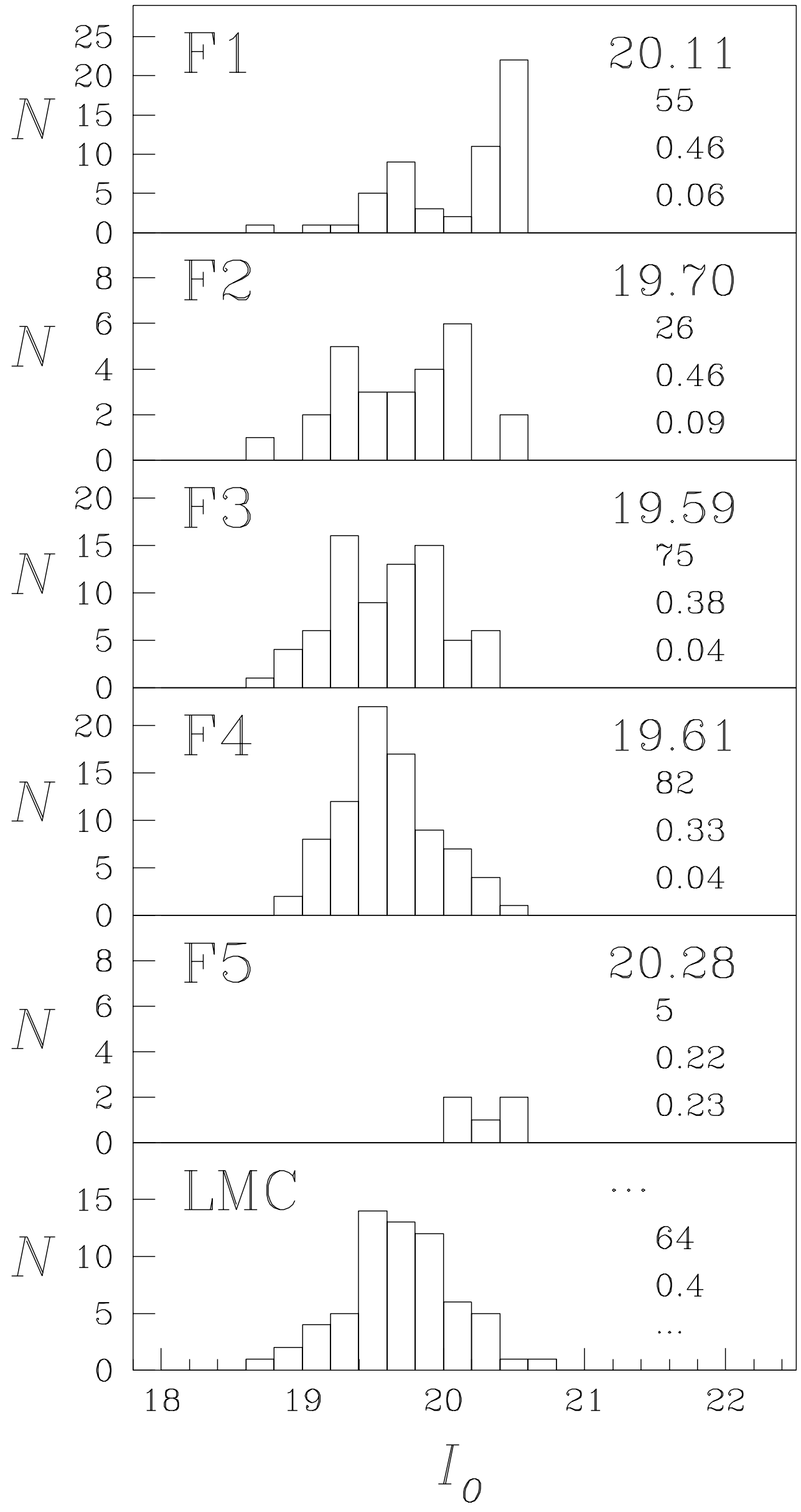

Fig. 12.— Carbon star luminosity functions for the five fields in M31, and that for the LMC Bar West field (Richer 1981). The C-stars were selected according to the color and magnitude criteria described in Sec. 4.1. The mean magnitude is given in the upper right hand corner of the panels, followed by the number of stars in the distribution, the standard deviation of the distribution and the error in the mean.

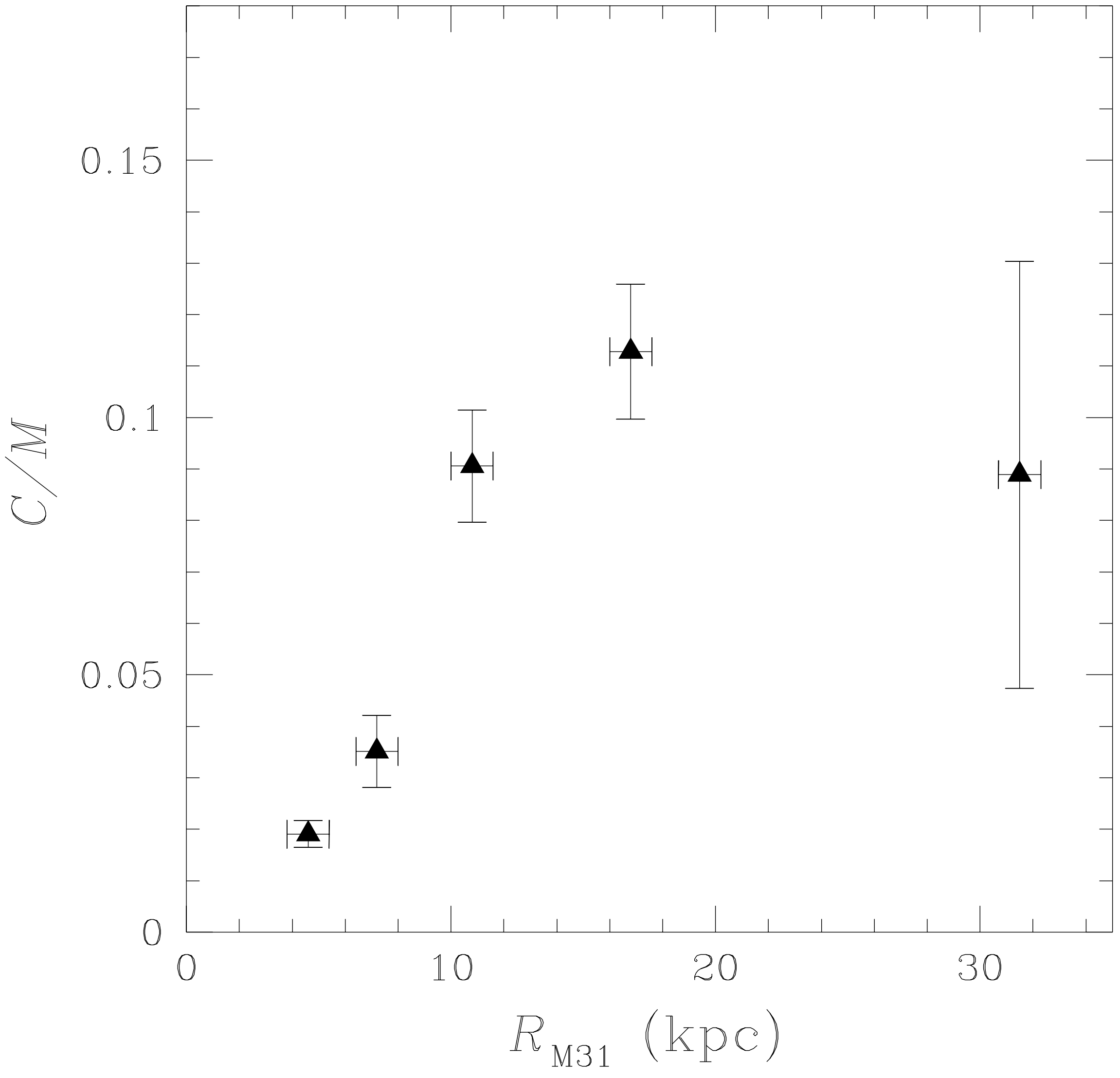

Fig. 13.— The measured C/M ratios in our five M31 fields as a function of $R_{\rm M31}$, with no completeness corrections. The C- and M-stars were selected according to the color and magnitude criteria described in Sec. 4.1. The vertical error bars show the Poissonian errors in the counts of the C- and M-stars, while horizontal error bars indicate the field size.

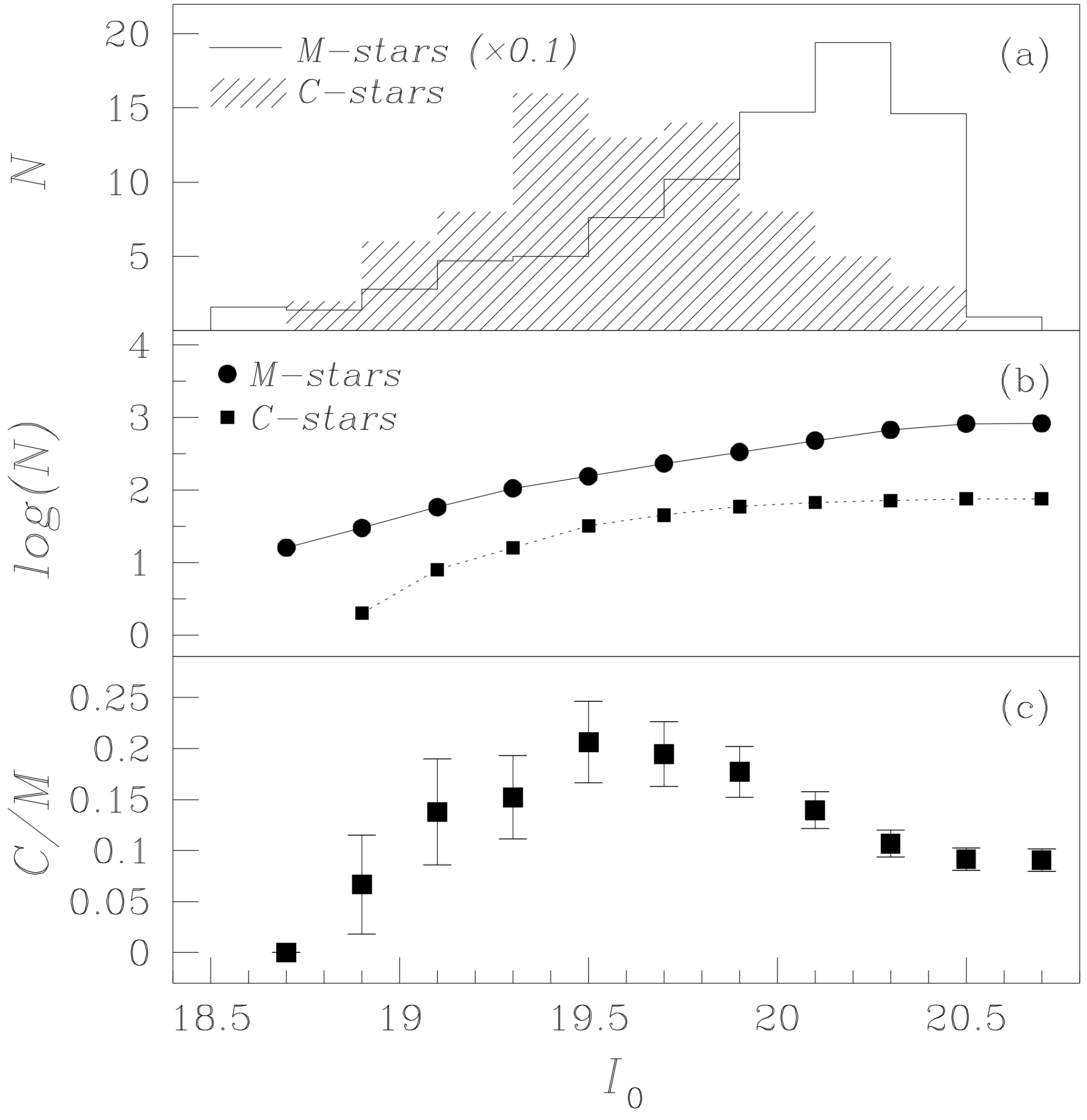

Fig. 14.— The upper panel shows the LFs of the C- and M-stars that were counted to derive the Field 3 C/M ratio. The M-star counts were multiplied by 0.1 to fit on the same scale as the C-stars. Panel (b) shows the logarithm of the cumulative counts of these stars with increasing magnitude. The final panel shows the measured C/M ratios derived when only those stars brighter than the abscissa magnitude are counted. The error bars in panel (c) are Poisson counting errors.

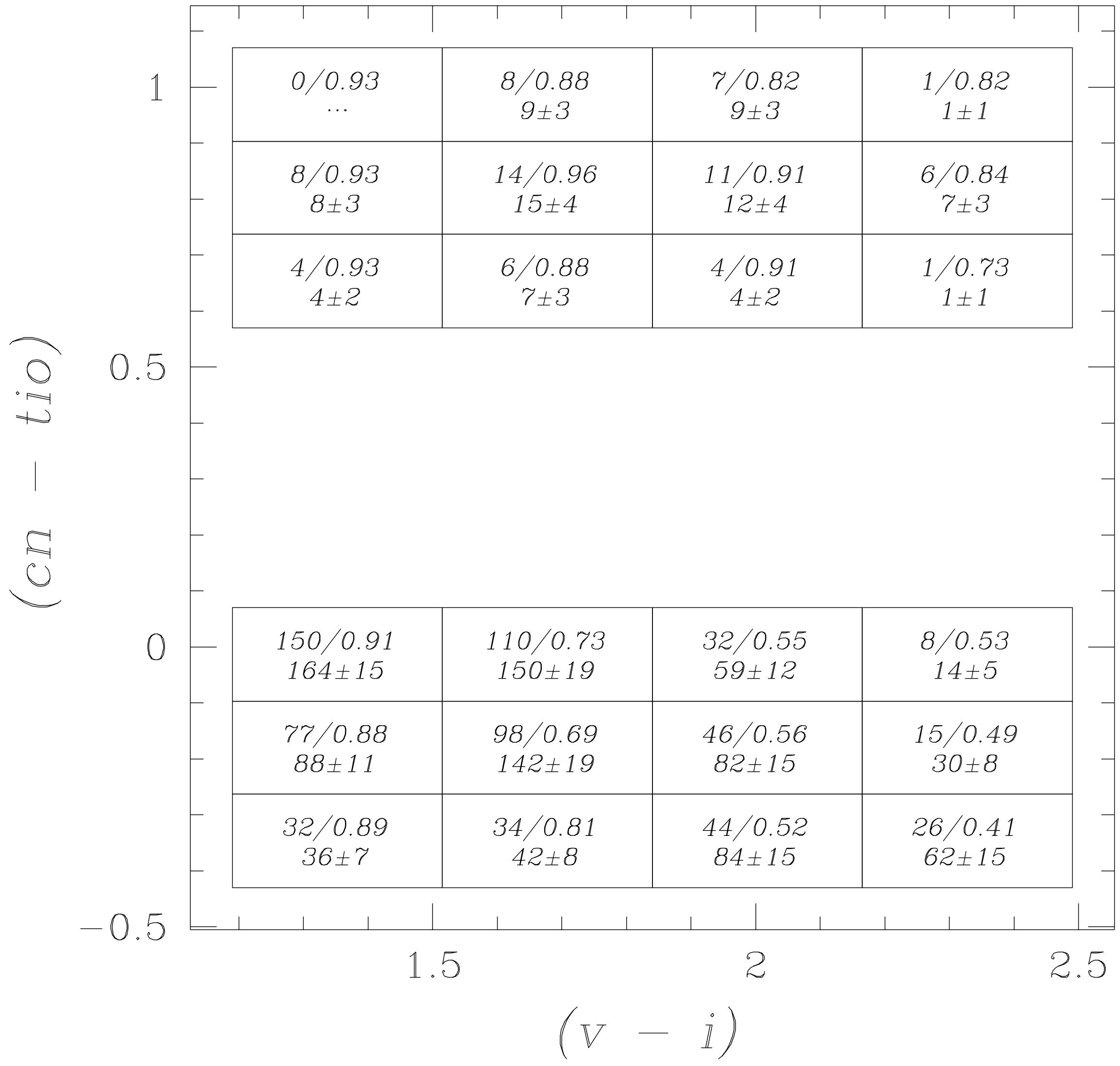

Fig. 15.— A diagram summarizing the results of the 2-color ADDSTAR tests in Field 3. The upper 12 bins in the $(cn-tio,\ v-i)$ plane cover the area in which a star is defined as a C-star, and the bottom 12 define the M-star region. The first number on the top row gives the number of real stars found in this bin together with incompleteness correction. The lower line gives the corrected counts along with their errors. An ellipsis indicates that no real stars were found in the bin.

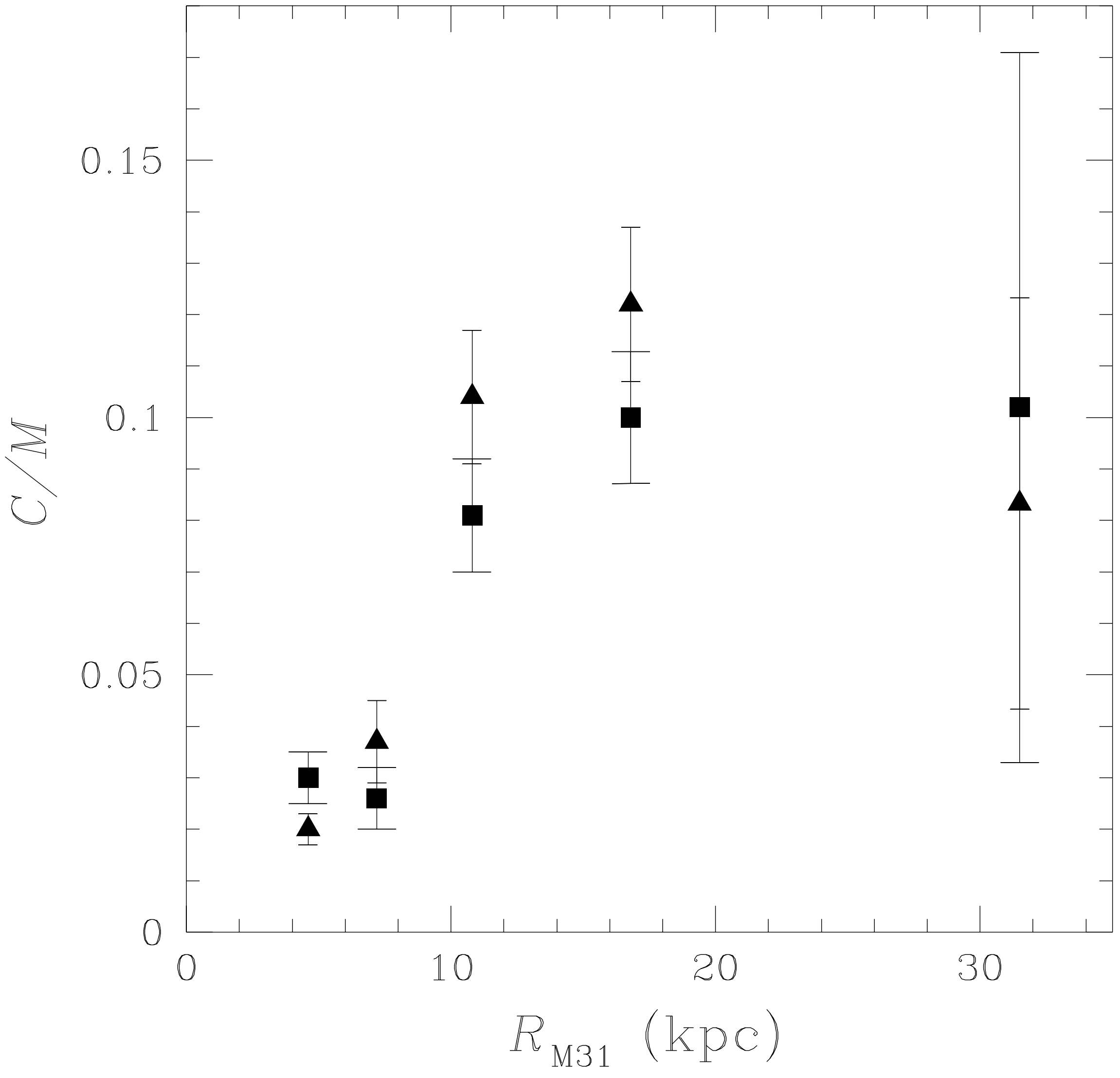

Fig. 16.— The measured C/M ratios in our five M31 fields as a function of $R_{M31}$. The C- and M-stars were selected according to the color and magnitude criteria described in Sec. 4.1, and the further color criteria given in Sec. 4.3.2. Triangles are uncorrected ratios, while squares indicate ratios corrected for incompleteness.

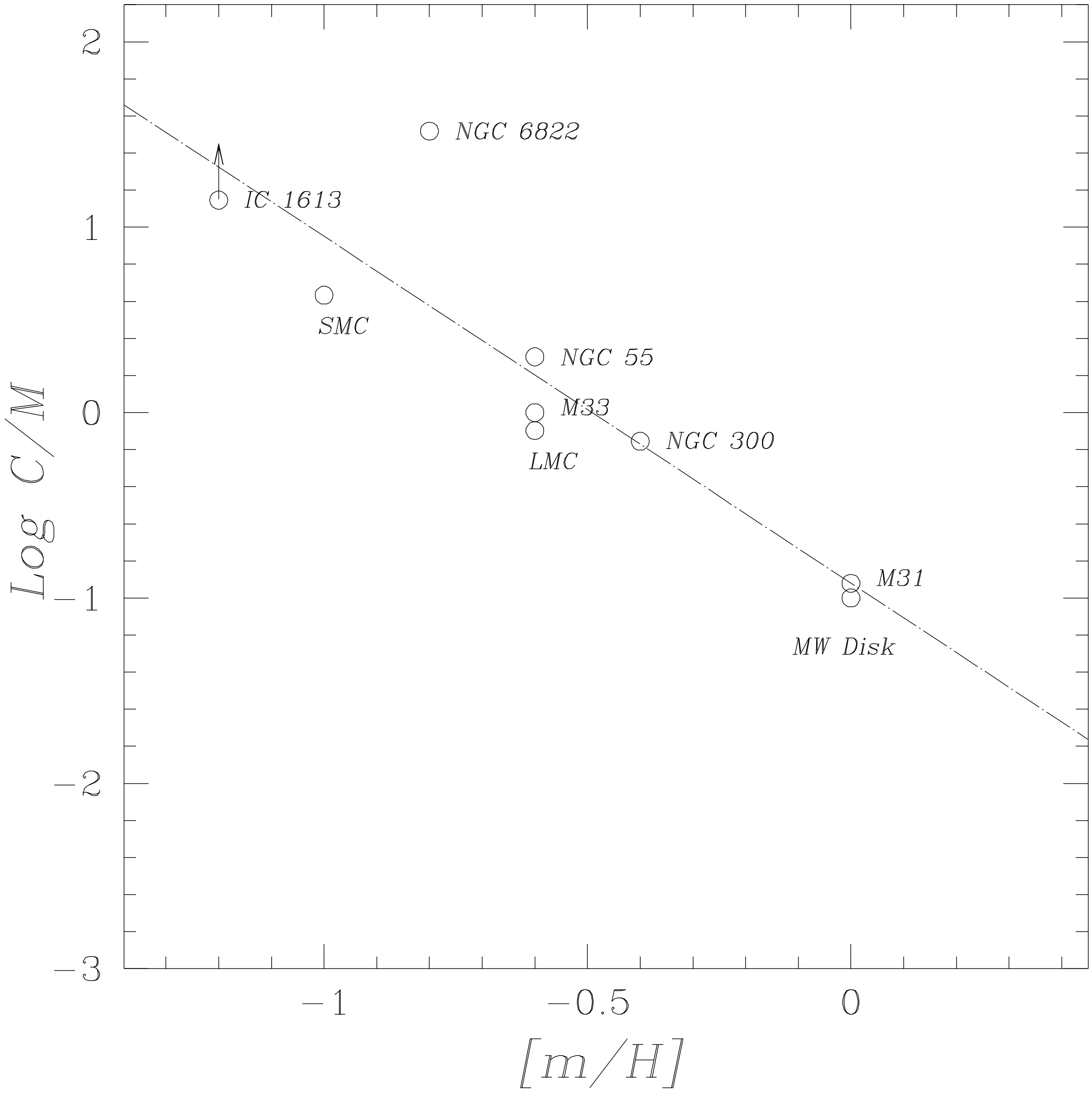

Fig. 17.— Open circles show data from Table 5 of Pritchet *et al.* (1987). The broken line shows the least-squares fit to the data used to derive (O/H) in Fig. 18.

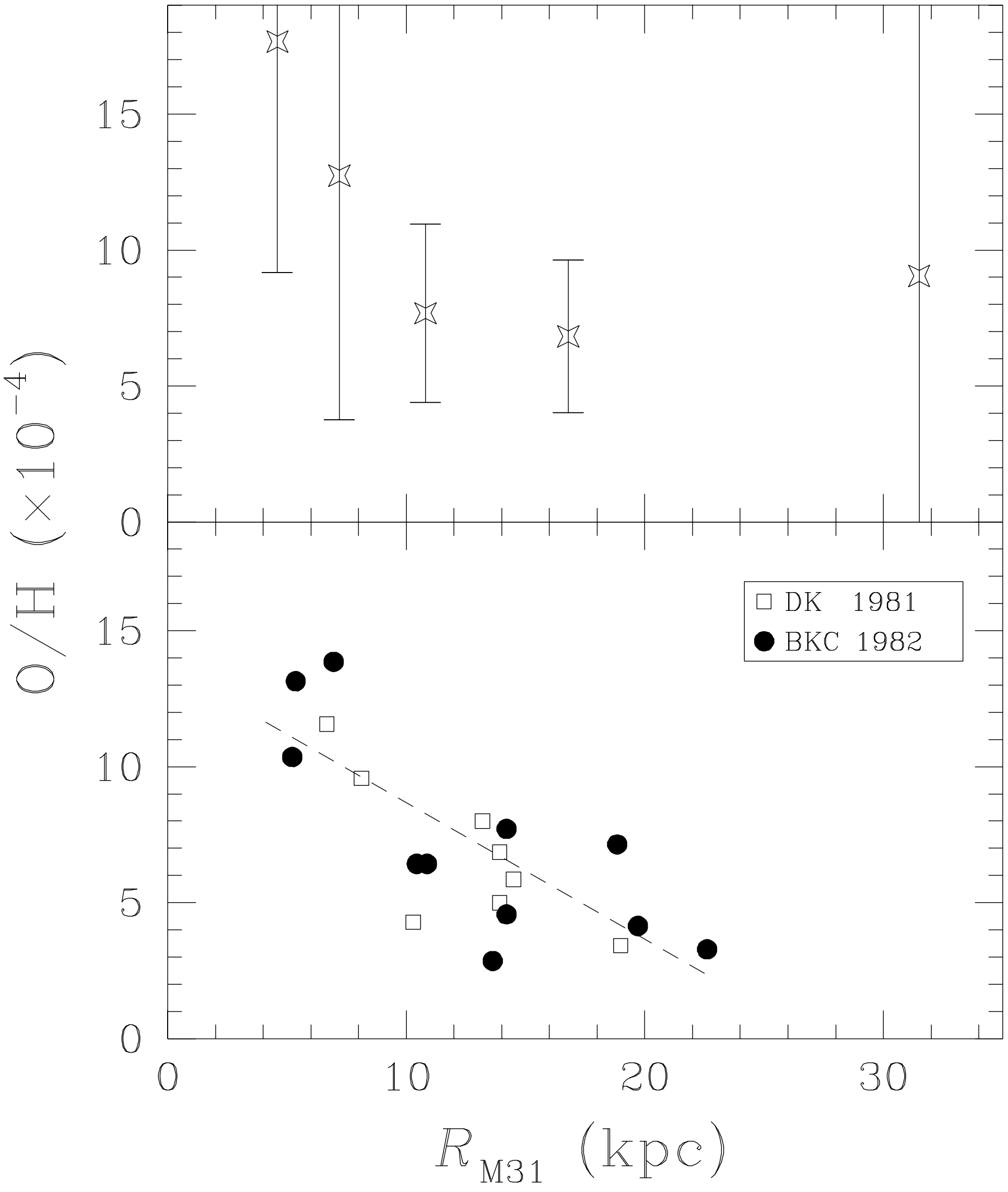

Fig. 18.— The data in the lower panel are from Fig. 6 of BKC82. The filled circles are measurements made by BKC82, while the open squares are data from Dennefeld & Kunth (1981). The dashed line is the fit of BKC82 to their (filled circles) data. The upper panel shows the number ratio of (O/H) derived from our C/M measurements. A full description of how these abundances were derived is given in the text. The errors shown in the upper panel were derived by propagation of the Poisson errors in the C/M ratio through Eqs. (3) and (4).

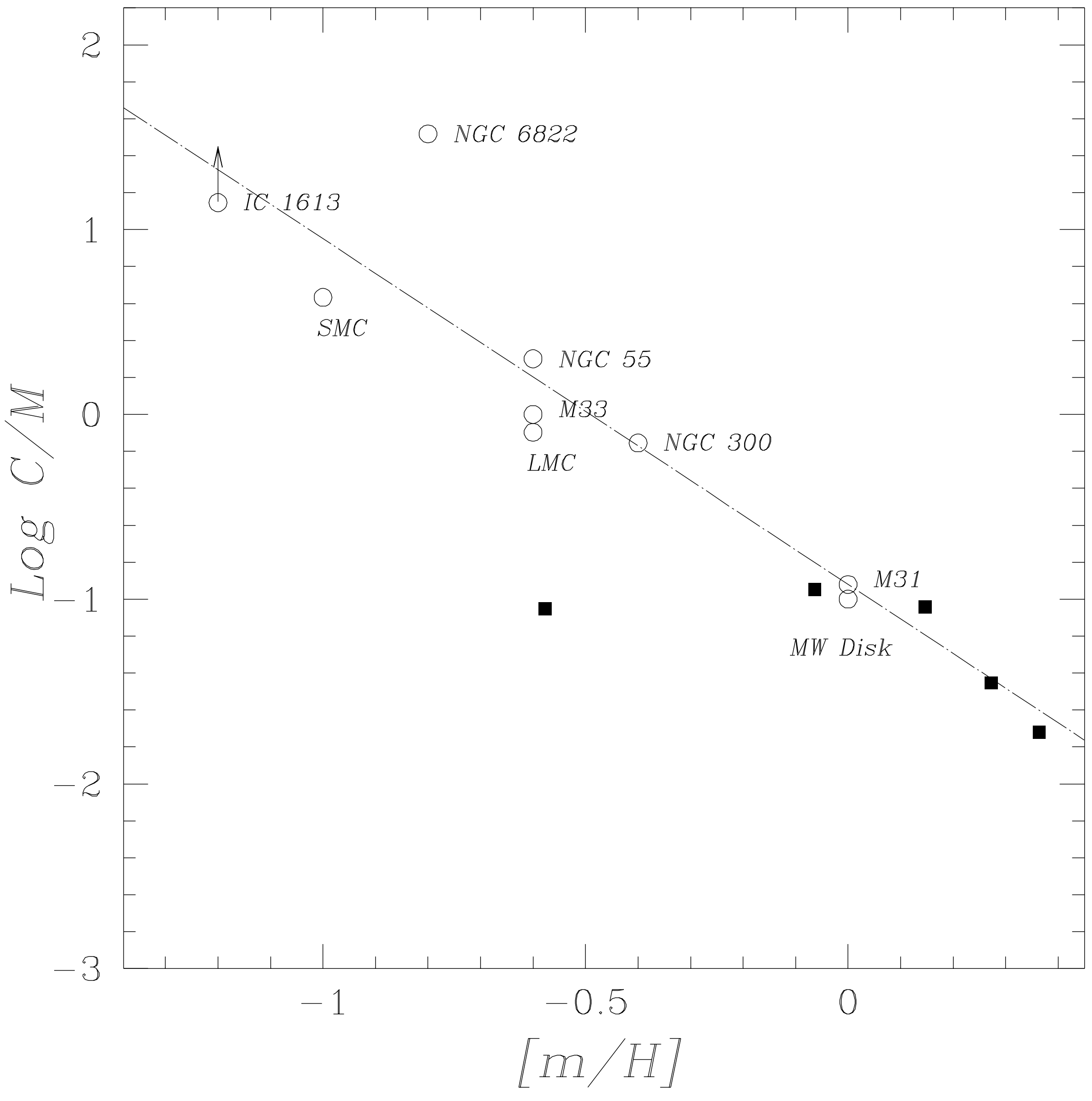

Fig. 19.— Open circles show data from Table 5 of Pritchet *et al.* (1987) and solid squares are data from this study. From left to right the solid squares are measurements from Fields 5 to 1, these have been placed using measurements from BKC82 and extrapolation of their relationship. The broken line shows the least-squares fit to the data (open circles) from Fig. 17.

Table 1: Observing Log for M31 Program Fields

| $\alpha_{2000}$, $\delta_{2000}$ | Filter | Exp (s) | FWHM | Fitted[a] | Night[b] |
|---|---|---|---|---|---|
| | | *Field 1* | | | |
| $\alpha = 0^{\rm h}41^{\rm m}41^{\rm s}$ | $CN$ | $3 \times 1200$ | $0\farcs78$ | 59742 | 1, 2[c] |
| $\delta = 40^\circ 59' 00''$ | $TiO$ | $3 \times 1200$ | $0\farcs78$ | 23909 | 1[c] |
| | $V$ | $3 \times 600$ | $0\farcs68$ | 40398 | 1, 4[c] |
| | $I$ | $3 \times 300$ | $0\farcs66$ | 158952 | 1, 4[c] |
| | | *Field 2* | | | |
| $\alpha = 0^{\rm h}41^{\rm m}08^{\rm s}$ | $CN$ | $3 \times 1200$ | $0\farcs82$ | 35626 | 3[c], 4[c] |
| $\delta = 40^\circ 50' 41''$ | $TiO$ | $3 \times 1200$ | $0\farcs76$ | 36427 | 3[c] |
| | $V$ | $4 \times 600$ | $0\farcs82$ | 9841 | 2[c], 4[c] |
| | $I$ | $4 \times 300$ | $0\farcs74$ | 69809 | 2[c], 3 |
| | | *Field 3* | | | |
| $\alpha = 0^{\rm h}40^{\rm m}22^{\rm s}$ | $CN$ | $3 \times 1200$ | $0\farcs78$ | 21141 | 3[c] |
| $\delta = 40^\circ 36' 24''$ | $TiO$ | $3 \times 1200$ | $0\farcs78$ | 21629 | 3[c] |
| | $V$ | $5 \times 600$ | $0\farcs80$ | 21737 | 2[c], 4[c] |
| | $I$ | $4 \times 300$ | $0\farcs78$ | 44935 | 2[c], 4[c] |
| | | *Field 4* | | | |
| $\alpha = 0^{\rm h}39^{\rm m}24^{\rm s}$ | $CN$ | $4 \times 1200$ | $0\farcs82$ | 15794 | 3[c] |
| $\delta = 40^\circ 14' 32''$ | $TiO$ | $3 \times 1200$ | $0\farcs76$ | 14830 | 3[c] |
| | $V$ | $2 \times 600$ | $0\farcs72$ | 8948 | 2, 4[c] |
| | $I$ | $3 \times 300$ | $0\farcs80$ | 14043 | 2[c] |
| | | *Field 5* | | | |
| $\alpha = 0^{\rm h}35^{\rm m}42^{\rm s}$ | $CN$ | $3 \times 1200$ | $0\farcs80$ | 1291 | 4[c] |
| $\delta = 39^\circ 29' 03''$ | $TiO$ | $3 \times 1200$ | $0\farcs66$ | 925 | 4[c] |
| | $V$ | $4 \times 600$ | $0\farcs76$ | 2226 | 4[c] |
| | $I$ | $4 \times 300$ | $0\farcs68$ | 950 | 2[c] |

[a]Number of stars fitted by ALLSTAR.

[b]1-4 refer to the nights of 27/09/91 to 30/09/91 respectively.

[c]Frames from these nights combined to make program frames.

Table 2: Serendipity.[†]

| Field | X,Y | What[a] | ID | Ref. |
|---|---|---|---|---|
| | | *Field 1* | | |
| 1 | 444, 316 | OC: | C204 | 1 |
| 1 | 648, 408 | GC | G117 | 1 |
| | | *Field 2* | | |
| 2 | 264, 624 | OC | C177 | 1 |
| 2 | 332, 1150 | GSC | ··· | 2 |
| 2 | 404, 876 | GSC | ··· | 2 |
| 2 | 1108, 508 | OC | C162 | 1 |
| 2 | 1882, 1874 | OC | C164 | 1 |
| | | *Field 3* | | |
| 3 | 655, 820 | OC | C107 | 1 |
| 3 | 694, 1280 | OC | C112 | 1 |

[†]Table 2 is presented in its complete form in the AAS CD-ROM Series, volume XX, 1995.

[a]OC: Open cluster.
GC: Globular cluster.
GSC: Star from HST Guide star catalog.
Ceph: Cepheid variable.
Irreg: Irregular variable.
Eclip: Eclipsing variable.
Short: Short period variable.

References. — (1) Hodge (1981); (2) Guide Star Catalog; (3) Baade & Swope (1965)

Table 3: Observing Log of Standard Fields Used for Calibration of M31 Data.

| Field | Airmass | Ref. |
|---|---|---|
| *Night 1 (27/09/91)* | | |
| SA110 364, 365 | 1.45 | 1 |
| NGC 7006 | 1.03, 1.4 | 2 |
| NGC 2419 | 1.3 | 2 |
| *Night 2 (28/09/91)* | | |
| SA111-1969 | 1.07 | 1 |
| SA110 | 1.08 | 1 |
| NGC 7006 | 1.03, 1.05, 1.35 | 2 |
| NGC 2419 | 1.27 | 2 |
| GD71 | 1.01 | 1 |

References. — (1) Landolt 1992; (2) Davis 1994

Table 4: Giant Branch Tip Determined by the Sobel Kernel.

| Field | 0.20 | 0.10 | 0.05 | Avg. |
|---|---|---|---|---|
| 1 | 20.7 | 20.65 | 20.725 | 20.69 |
| 2 | 20.7 | 20.75 | 20.725 | 20.73 |
| 3 | 20.7 | 20.95 | 21.175 | 20.94 |
| 4 | 20.7 | 20.75 | 20.975 | 20.81 |
| 5 | 20.9 | 20.95 | 21.225 | 21.03 |

Table 5: Observed, 'Photometric' and Intrinsic Widths of Giant Branches

| Field | $\sigma_{obs}$ | $\sigma_{phot}$ | $\sigma_{int}$ | FWHM$_{int}$ |
|---|---|---|---|---|
| 1 | 0.35 | 0.28 | 0.21 | 0.49 |
| 2 | 0.36 | 0.25 | 0.26 | 0.61 |
| 3 | 0.47 | 0.22 | 0.42 | 0.99 |
| 4 | 0.25 | 0.15 | 0.20 | 0.47 |
| 5 | 0.24 | 0.11 | 0.21 | 0.49 |

Table 6: C- and M-star Counts in M31.

| Field | $R_{\rm M31}$ | C[a] | M[a] | C[b] | M[b] |
|---|---|---|---|---|---|
| 1 | 4.6 kpc | 108 | 4514 | 55 | 2884 |
| 2 | 7.2 kpc | 29 | 1092 | 26 | 740 |
| 3 | 10.8 kpc | 75 | 1334 | 75 | 828 |
| 4 | 16.8 kpc | 82 | 1225 | 82 | 727 |
| 5 | 31.5 kpc | 14 | 108 | 5 | 75 |

[a]Color criteria only

[b]Color and magnitude criteria.

Table 7: Coordinates and Photometry of Candidate C-Stars.[†]

| ID | X,Y | $I$ | $(V-I)$ | $(CN-TiO)$ |
|---|---|---|---|---|
| | | *Field 1* | | |
| f1C.1 | 153, 898 | 19.15(0.02) | 2.42(0.03) | 0.62(0.03) |
| f1C.2 | 564, 240 | 19.43(0.01) | 3.15(0.03) | 0.38(0.04) |
| f1C.3 | 823, 1132 | 19.63(0.02) | 2.90(0.03) | 0.34(0.03) |
| f1C.4 | 250, 1146 | 19.83(0.03) | 5.08(0.38) | 0.65(0.05) |
| f1C.5 | 1272, 605 | 19.87(0.04) | 3.92(0.14) | 0.34(0.05) |
| f1C.6 | 578, 153 | 19.98(0.02) | 3.55(0.07) | 1.04(0.04) |
| f1C.7 | 943, 322 | 19.99(0.03) | 2.52(0.05) | 0.59(0.05) |
| f1C.8 | 322, 1994 | 20.00(0.02) | 2.81(0.04) | 0.57(0.04) |
| f1C.9 | 72, 1330 | 20.04(0.02) | 2.56(0.04) | 0.32(0.05) |
| f1C.10 | 1343, 1038 | 20.06(0.02) | 3.02(0.06) | 0.61(0.05) |

[†]Table 7 is presented in its complete form in the AAS CD-ROM Series, volume XX, 1995.

Table 8: Coordinates and Photometry of Candidate M-Stars.[†]

| ID | X,Y | $I$ | $(V-I)$ | $(CN-TiO)$ |
|---|---|---|---|---|
| | | *Field 1* | | |
| f1M.1 | 705, 862 | 18.95(0.02) | 3.79(0.05) | –1.13(0.04) |
| f1M.2 | 1816, 421 | 19.04(0.02) | 2.62(0.06) | –0.46(0.02) |
| f1M.3 | 1419, 1151 | 19.06(0.02) | 2.26(0.02) | –0.51(0.04) |
| f1M.4 | 1517, 508 | 19.07(0.03) | 3.33(0.05) | –0.50(0.04) |
| f1M.5 | 576, 1667 | 19.11(0.03) | 3.86(0.05) | –1.37(0.04) |
| f1M.6 | 309, 798 | 19.21(0.01) | 2.59(0.03) | –0.24(0.03) |
| f1M.7 | 823, 1210 | 19.23(0.02) | 2.38(0.03) | –0.40(0.03) |
| f1M.8 | 1252, 1682 | 19.25(0.01) | 3.44(0.04) | –1.17(0.03) |
| f1M.9 | 447, 1772 | 19.26(0.01) | 3.13(0.05) | –0.83(0.03) |
| f1M.10 | 693, 1973 | 19.26(0.03) | 2.92(0.06) | –0.45(0.04) |

[†]Table 8 is presented in its complete form in the AAS CD-ROM Series, volume XX, 1995.